\documentclass[11pt,a4paper]{article}
\usepackage{jcappub}

\usepackage{bm}
\usepackage{dsfont}
\usepackage{bbm}
\usepackage{color}

\allowdisplaybreaks[1]

\newcommand{\ii}{\mathbbm{i}}
\newcommand{\di}{\! \text{d}}
\newcommand{\dd}{\text{d}}
\newcommand{\ee}{\text{e}}
\renewcommand\({\left(}
\renewcommand\){\right)}
\renewcommand\[{\left[}
\renewcommand\]{\right]}
\newcommand{\CLASS}{\texttt{CLASS}}
\newcommand{\CAMB}{\texttt{CAMB}}
\newcommand{\diff}[2]{\frac{\dd #1}{\dd #2}}
\newcommand{\diffpar}[2]{\frac{\partial #1}{\partial #2}}
\newcommand{\I}{\mathds{1}}
\newcommand{\kappadot}{\dot{\kappa}}
\newcommand{\nhat}{\hat{n}}
\newcommand{\Huphase}{\ee^{\ii \delta(\vec{x},\vec{k})}}

\newcommand{\spY}[3]{{}_{#1}^{\vphantom{#3}}Y_{#2}^{#3}}
\newcommand{\spE}{\mathcal{E}}
\newcommand{\spB}{\mathcal{B}}

\newcommand{\Yttpr}{Y_2^{2\prime}}

\newcommand{\Ytepr}{Y_2^{1\prime}}

\newcommand{\Ytn}{Y_2^0}

\newcommand{\Ytnpr}{Y_2^{0\prime}}

\newcommand{\Ytm}{Y_2^m}
\newcommand{\pYtm}{\spY{2}{2}{m}}
\newcommand{\mYtm}{\spY{-2}{2}{m}}
\newcommand{\Ytmpr}{Y_2^{m\prime}}
\newcommand{\pYtmpr}{\spY{2}{2}{m\prime}}
\newcommand{\mYtmpr}{\spY{-2}{2}{m\prime}}
\newcommand{\phYtm}{\spY{\phantom{-}2}{2}{m}}

\newcommand{\spK}[3]{{}_{#1}^{\vphantom{#3}}\kappa_{#2}^{#3}}

\newcommand{\Wthreej}[6]{\begin{pmatrix} #1 & #3 & #5 \\ #2 & #4 & #6 \end{pmatrix}}
\newcommand{\sourceF}{\mathfrak{u}}
\newcommand{\sourceG}{\mathfrak{v}}

\newcommand{\vrarray}{\rule[-0.3cm]{0cm}{1cm}}
\newcommand{\vrsarray}{\rule[-0.15cm]{0cm}{0.7cm}}
\newcommand{\spM}[3]{{}_{#1}^{\vphantom{#3}}M_{#2}^{#3}}
\newcommand{\cotk}{\text{cot}_\text{K}}

\newcommand{\lmax}{l_\text{max}}

\begin{document}

\hfill CERN-PH-TH/2013-099, LAPTH-025/13

\title{Optimal polarisation equations in FLRW universes}
\author[a]{Thomas Tram}
\author[a,b,c]{and Julien Lesgourgues}
\affiliation[a]{Institut de
Th\'eorie des Ph\'enom\`enes Physiques, \'Ecole Polytechnique
F\'ed\'erale de Lausanne, CH-1015, Lausanne,
Switzerland}
\affiliation[b]{CERN, Theory Division, CH-1211 Geneva 23, Switzerland}
\affiliation[c]{LAPTh (CNRS - Universit\'e de Savoie), BP 110, F-74941 Annecy-le-Vieux Cedex, France}
\emailAdd{thomas.tram@epfl.ch}
\emailAdd{Julien.Lesgourgues@cern.ch}

\abstract{This paper presents the linearised Boltzmann equation for photons for scalar, vector and tensor perturbations in flat, open and closed FLRW cosmologies. We show that E- and B-mode polarisation for all types can be computed using only a single hierarchy. This was previously shown explicitly for tensor modes in flat cosmologies but not for vectors, and not for non-flat cosmologies.}
\maketitle

\section{Prelude}

\subsection{Introduction}
The Boltzmann equation for linear cosmological perturbations constitutes a set of roughly one hundred coupled, ordinary differential equations (ODE), depending on the assumed cosmological model and the requested accuracy. This system of ODE's are solved several thousand times a day, which easily makes it the most often solved system of differential equations in cosmology. The equations were derived in a series of papers with contributions from many authors. The effect of photon polarisation for scalar perturbations was first correctly included by Kaiser~\cite{Kaiser:1983} and Bond\&Efstathiou~\cite{Bond:1984fp}. The equations for tensor perturbations were given by Crittenden et. al.~\cite{Crittenden:1993ni} building on previous work by Polnarev~\cite{Polnarev:1985}. Kosowsky~\cite{Kosowsky:1994cy} gave a quantum mechanical re-derivation of the scalar and tensor photon Boltzmann equation. Seljak\&Zaldarriaga invented the line-of-sight method~\cite{Seljak:1996is} for calculating CMB-anisotropies which drastically reduced the number of differential equations necessary for calculating the power-spectrum of anisotropies. They were also the first to realise that the Stokes parameters $Q$ and $U$ were not optimal for an all-sky analysis, but that the parity eigenstates $E$ and $B$ should be used instead~\cite{Zaldarriaga:1996xe}. 

Hu\&White finally gave a unified treatment of all modes in their seminal paper~\cite{Hu:1997hp}. Their approach is called the total angular momentum method, and it is based upon expansions in spin-weighted spherical harmonics. In this formalism $E_l$ and $B_l$ are the natural expansion coefficients in a spin-weighted expansion of $Q\pm\ii U$. Due to the relation between spin-weighted spherical harmonics and rotation matrices, this formulation gave a clean derivation of the general scattering term in the photon Boltzmann equation. At the same time, Seljak, Zaldarriaga and Bertschinger had derived the scalar equations in non-flat cosmologies~\cite{Zaldarriaga:1997va}. Hu\&White, in collaboration with Seljak and Zaldarriaga, finally extended the total angular momentum method to non-flat cosmologies~\cite{Hu:1997mn}, thereby giving equations valid for all modes in cosmologies with constant curvature. However, the reduction from 2 polarisation hierarchies to 1 for tensor modes, which was previously used in~\cite{Polnarev:1985,Kosowsky:1994cy,Crittenden:1993ni,Zaldarriaga:1996xe} was lost in this process. These equations are also not easy to compare with the ones of Ma\&Bertschinger~\cite{Ma:1995ey} which for many cosmologists continue to be the primary reference on cosmological perturbation theory. Thus, the purpose of this paper is to show how the number of polarisation hierarchies can be reduced in all cases, and also to connect the modern paper of Hu et. al. to the classic paper of Ma\&Bertschinger. 

There is another formalism that we did not mention so far, namely the covariant and gauge invariant approach of Challinor\&Lasenby, see~\cite{Challinor:1998xk} and references herein. This approach was further expanded and refined by Challinor~\cite{Challinor:1999xz,Challinor:2000as} and later also by Lewis~\cite{Lewis:2002nc}. This approach shares many of the advantages of the total angular momentum method, and it can be easily generalised to non-linear perturbations. The equations of the popular Boltzmann code \CAMB{} is derived in this formalism~\cite{Lewis:2004ef}. However, this method also shares the disadvantage of requiring two polarisation hierarchies for vectors and tensors.

\subsection{Conventions}
We are using the $({-}{+}{+}{+})$ sign-convention for the metric, and greek indices are running from $0$ to $3$ while latin indices are running from $1$ to $3$. For consistency with Hu et al.~\cite{Hu:1997hp}, we are omitting the Condon-–Shortley phase $(-1)^m$ in the definition of the (spin-weighted) spherical harmonics and in the definition of the associated Legendre polynomials. Note that this is contrary to the conventions of both Wikipedia and Mathematica, which includes the phase for both. Under this convention, the following equations hold:
\begin{align}
P_l^m(\mu) &= (1-\mu^2)^{m/2} \diff{^m}{\mu^m}\(P_l(\mu)\), \label{eq:Plmdef} \\
Y_l^m(\theta,\phi) &= \sqrt{\frac{2l+1}{4\pi} \frac{(l-m)!}{(l+m)!}} P_l^m(\mu) \ee^{\ii m \phi}, \label{eq:Ylmdef}
\end{align}
where $\mu\equiv \cos \theta$.
Throughout this paper we will not write the arguments of these functions explicitly. Instead we employ the following notation:
\begin{align}
\spY{s}{l}{m} &\equiv \spY{s}{l}{m} (\theta, \phi) \\
\spY{s}{l}{m\prime} &\equiv \(\spY{s}{l}{m}(\theta',\phi') \)^* \\
P_l^m &\equiv P_l^m(\mu) \\
P_l^{m\prime} &\equiv P_l^{m\prime}(\mu')
\end{align}
where the conventions for ordinary spherical harmonics and ordinary Legendre polynomials follows by putting $s=0$ and $m=0$ respectively. Note the implicit complex conjugation for $Y_l^{m\prime}$ and $\spY{s}{l}{m\prime}$. We will use Ma\&Bertschinger's convention for Legendre expansions
\begin{equation}
 X(\mu) = \sum_{l=0}^\infty (-\ii)^l (2l+1) X_l P_l, \label{eq:legexpansion}
\end{equation}
and it will be understood that $X' \equiv X(\mu')$.

\subsection{Metric}
Following~\cite{Hu:1997mn}, we write the metric as
\begin{equation}
g_{\mu\nu} = a^2 \(\gamma_{\mu\nu} + h_{\mu\nu} \),
\end{equation}
where $h_{\mu\nu}$ is a perturbation and the spatial part of the background metric can be written as
\begin{equation}
\gamma_{ij} = \frac{1}{|K|} \( d\xi^2 +  \begin{pmatrix} \sinh^2 \xi \\ \xi^2 \\ \sin^2 \xi \end{pmatrix} (d\theta^2 + \sin^2\theta d\phi^2)  \), \begin{matrix} K<0 \\ K\rightarrow 0 \\K>0 \end{matrix}.
\end{equation}
This metric can be constructed by embedding a 3-dimensional sphere (pseudo-sphere) of positive (negative) curvature $K$ in 4 dimensional Euclidean (Lorentzian) space. The radial coordinate $r$ of the 4-dimensional space is then mapped to $\xi$ by
\begin{equation}
\xi = \left\{ \begin{array}{ll} \text{arcsinh}(\sqrt{|K|}r), & K<0, \\
\sqrt{|K|}r, & K\rightarrow 0, \\
\arcsin (\sqrt{|K|}r), & K>0. \end{array} \right.
\end{equation}
The covariant derivative of $X$ with respect to the spatial background metric $\gamma_{ij}$ will be denoted $X_{|i}$. We will use conformal time $\tau$, and the derivative with respect to $\tau$ is denoted by a dot. $K$ is constant and given by $K=-H_0^2(1-\Omega_\text{tot})$.

\subsection{Perturbation types and normal modes}
Cosmological perturbations are usually divided into three types: scalar-, vector- and tensor-perturbations. This decomposition is based on how the perturbation behaves under spatial rotations, and it is useful because the Boltzmann equation does not couple the different types at linear order. More formally, we can write the eigentensor equation for the Laplacian:
\begin{equation}
\nabla^2 \mathcal{Q}^{(m)}_{i_1i_2\cdots i_{|m|}} = \gamma^{jk} \mathcal{Q}^{(m)}_{i_1i_2\cdots i_{|m|}|jk} = -k^2 \mathcal{Q}^{(m)}_{i_1i_2\cdots i_{|m|}}. \label{eq:Qeigen}
\end{equation}
The vector modes has zero divergence, i.e. $\gamma^{ij} \mathcal{Q}_{i|j}^{(\pm 1)}=0$ while the tensor modes are transverse and traceless: $\gamma^{ik} \mathcal{Q}_{ij|k}^{(\pm 2)} = \gamma^{ij} \mathcal{Q}_{ij}^{(\pm 2)}=0$. These eigentensors, together with the three auxiliary tensors
\begin{equation}\label{eq:Qaux}
\mathcal{Q}_i^{(0)}=-\frac{1}{k} \mathcal{Q}^{(0)}_{|i}, \quad 
\mathcal{Q}_{ij}^{(0)} = \frac{1}{k^2} \mathcal{Q}_{|ij}^{(0)} + \frac{1}{3} \gamma_{ij} \mathcal{Q}^{(0)}, \quad 
\mathcal{Q}_{ij}^{(\pm 1)} = -\frac{1}{2k} \(\mathcal{Q}_{i|j}^{(\pm 1)} + \mathcal{Q}_{j|i}^{(\pm 1)} \), 
\end{equation}
can be used to decompose a general perturbation such as the perturbed metric or the baryon velocity. In flat space, these objects have simple explicit representations~\cite{Hu:1997hp}. In the Boltzmann equation, all $\mathcal{Q}^{(m)}$-tensors are fully contracted with the propagation unit vector for the photons, $\hat{n}$, or with $(\hat{e}_1 \pm \ii \hat{e}_2)$ in the case of polarisation\footnote{$\hat{e}_1$ and $\hat{e}_2$ form an orthonormal basis in the plane perpendicular to the wave vector $\vec{k}$.}. These fully contracted $\mathcal{Q}^{(m)}$-tensors are called normal modes and can be used to expand any function of any type $m$ and spin $s$. In flat space, the $s=0$ and $s\pm 2$ normal modes can be written as
\begin{align}
M_l^m &= (-\ii)^l \sqrt{\frac{4\pi}{2l+1}} Y_l^m \ee^{\ii \vec{k} \cdot \vec{x}}, &
\spM{\pm2}{l}{m} &=  (-\ii)^l \sqrt{\frac{4\pi}{2l+1}} \spY{\pm2}{l}{m} \ee^{\ii \vec{k} \cdot \vec{x}},  \label{eq:Mlm}
\end{align}
and for a general spin $s$ we have
\begin{align}
\spM{s}{l}{m} &=  (-\ii)^l \sqrt{\frac{4\pi}{2l+1}} \spY{s}{l}{m} \ee^{\ii \vec{k} \cdot \vec{x}}.
\end{align}
We always assume that the direction of propagation $\hat{n}$ is expressed in spherical coordinates $(\theta,\phi)$ where $\theta$ is the angle between $\hat{n}$ and the wave vector $\vec{k}$ of each Fourier mode. It was a key insight by Hu et. al. to realise that one can construct modes with the same angular structure in a non-flat space. They found that the normal modes can be written in the form
\begin{align}
\spM{s}{l}{m} &=  (-\ii)^l \sqrt{\frac{4\pi}{2l+1}} \spY{s}{l}{m} \Huphase ,
\end{align}
where $\Huphase$ can in principle be calculated, along with a general formula for the spatial derivative of a mode:
\begin{align}\label{eq:Mderivative}
n^i \(\spM{s}{l}{m} \)_{|i} &=\frac{q}{2l+1} \spK{s}{l}{m} \(\spM{s}{l-1}{m} \) - \spK{s}{l+1}{m} \( \spM{s}{l+1}{m} \) -\ii \frac{q m s}{l+1} \spM{s}{l}{m},
\end{align}
\begin{flalign*}
&\text{where} \quad \spK{s}{l}{m} = \sqrt{\frac{(l^2-m^2)(l-s^2)}{l^2}\(1-K \frac{l^2}{q^2}\)} \quad \text{and} \quad 
q \equiv \sqrt{|K|} \nu = \sqrt{k^2+(|m|+1)K}.&
\end{flalign*}
We shall also define the wavevector $\vec{q}$ which is parallel to $\vec{k}$ and has length $q$. Note that equation~\eqref{eq:Mderivative} tells us everything we need to know about the generalised plane wave $\Huphase$: it will just cancel out at each side of the equations just like the usual plane wave. Because of this, we will almost never write it explicitly in our equations. 

\section{Boltzmann equation}
\subsection{Temperature and polarisation}
The CMB radiation can be described by 3 Stokes parameters $\mathcal{I}$, $\mathcal{Q}$ and $\mathcal{U}$. The 4th Stokes parameter $\mathcal{V}$ represents circular polarisation but it is irrelevant since it is not generated by Thomson scattering. Following Hu\&White~\cite{Hu:1997hp}, we form the vector of first order perturbations\footnote{Two different conventions exist for writing the perturbations, which differ by a factor 4. The first convention is using intensity fluctuation $(\delta \mathcal{I}/T, \ldots)$ and is being used by Ma\&Bertschinger~\cite{Ma:1995ey}, Crittenden et. al.~\cite{Crittenden:1993ni} and Kosowsky~\cite{Kosowsky:1994cy}. The second convention uses the equivalent of temperature fluctuations $(\Theta,\ldots)$ and is being used by Hu\&White~\cite{Hu:1997mn,Hu:1997hp}. Seljak\&Zaldarriaga~\cite{Seljak:1996is,Zaldarriaga:1996xe,Zaldarriaga:1997va} are using the second convention for tensors and the first convention for scalars. We will make use of both.}
\begin{equation}
\vec{T}(\vec{x},\nhat,\tau) = \frac{1}{4} \(\frac{\delta \mathcal{I}}{T}, \frac{\delta \mathcal{Q}+\ii \delta \mathcal{U}}{T}, \frac{\delta \mathcal{Q}-\ii \delta \mathcal{U}}{T} \) \equiv \(\Theta, Q+\ii U, Q-\ii U \), \label{eq:Boltzmann}
\end{equation}
where $\nhat$ is the propagation direction of the photons and $\Theta$ is the temperature fluctuation $\Theta \equiv \delta T/T$. We start from the Boltzmann equation for $\vec{T}$, equation~(45) in Hu\&White~\cite{Hu:1997hp}
\begin{equation}
\diff{\vec{T}}{\tau} = \diffpar{\vec{T}}{\tau} + n^i \nabla_i \vec{T} = \vec{C}[\vec{T}] + \vec{D}[h_{\mu\nu}],
\end{equation}
where $\vec{D}=(D_\Theta,0,0)$ is the source term related to the metric, and the collision term can be written as
\begin{equation}
\vec{C}[\vec{T}] = -\kappadot \vec{T} +\kappadot \(\int \frac{\di\Omega'}{4\pi} \Theta' + \nhat \cdot \vec{v}_B,0,0\)
+\frac{\kappadot}{10} \int \di \Omega' \sum_{m=-2}^{2} \mathbf{P}^{(m)}\( \Omega, \Omega' \) \vec{T}'.
\end{equation}
Here $\vec{v}_B$ is the baryon velocity, $\kappadot=n_e \sigma_T a$ is the differential optical depth, and the scattering matrix can be written in the following form:
\begin{equation}
\mathbf{P}^{(m)}\( \Omega, \Omega' \) = 
\begin{pmatrix}
\Ytmpr\Ytm & -\sqrt{\frac{3}{2}} \pYtmpr \Ytm & -\sqrt{\frac{3}{2}} \mYtmpr \Ytm \\
-\sqrt{6} \Ytmpr \phYtm & 3\pYtmpr \phYtm & 3\mYtmpr \phYtm \\
-\sqrt{6} \Ytmpr \mYtm  & 3\pYtmpr \mYtm  & 3\mYtmpr \mYtm
\end{pmatrix}.
\end{equation}
Evaluating the matrix product yields:
\begin{align*}
\mathbf{P}^{(m)} \vec{T}' &=
\begin{pmatrix}
\Ytm \left\{\Ytmpr \Theta' -\sqrt{\frac{3}{2}} \pYtmpr (Q' + \ii U') -\sqrt{\frac{3}{2}} \mYtmpr (Q' - \ii U') \right\} \\
\spY{\phantom{-}2}{2}{m} \left\{-\sqrt{6} \Ytmpr \Theta +3 \pYtmpr (Q' + \ii U') +3 \mYtmpr (Q' - \ii U') \right\} \\
\mYtm \left\{-\sqrt{6} \Ytmpr \Theta +3 \pYtmpr (Q' + \ii U') +3 \mYtmpr (Q' - \ii U') \right\}
\end{pmatrix}
\\
&=
\begin{pmatrix}
\Ytm \left\{\Ytmpr \Theta' -\sqrt{\frac{3}{2}} (\pYtmpr+\mYtmpr) Q' -\sqrt{\frac{3}{2}} (\pYtmpr-\mYtmpr) \ii U' \right\} \\
\spY{\phantom{-}2}{2}{m} \left\{-\sqrt{6} \Ytmpr \Theta +3 (\pYtmpr+\mYtmpr) Q' + 3(\pYtmpr-\mYtmpr) \ii U' \right\} \\
\mYtm \left\{-\sqrt{6} \Ytmpr \Theta +3 (\pYtmpr+\mYtmpr) Q' + 3(\pYtmpr-\mYtmpr) \ii U' \right\}
\end{pmatrix} \\
&=
\begin{pmatrix}
\Ytm \left\{\Ytmpr \Theta' -\sqrt{\frac{3}{2}} \spE^{m\prime} Q' -\sqrt{\frac{3}{2}} \spB^{m\prime} \ii U' \right\} \\
\spY{\phantom{-}2}{2}{m} \left\{-\sqrt{6} \Ytmpr \Theta +3 \spE^{m\prime} Q' + 3\spB^{m\prime} \ii U' \right\} \\
\mYtm \left\{-\sqrt{6} \Ytmpr \Theta +3 \spE^{m\prime} Q' + 3\spB^{m\prime} \ii U' \right\}
\end{pmatrix}.
\end{align*}
where we defined the symbols
\begin{align}
\spE^m &\equiv \pYtm+\mYtm, &\spE^{m\prime} \equiv \pYtmpr+\mYtmpr, \\
\spB^m &\equiv \pYtm-\mYtm, &\spB^{m\prime} \equiv \pYtmpr-\mYtmpr.
\end{align}
The explicit representations of these symbols for $m=0,1\text{ and }2$ are given in table~\ref{tab:EandB}.
\begin{table}%
\begin{center}
\begin{tabular}{l|ccc}
$m$    & $Y_2^m$ &                             $\spE^m$                              &$\spB^m$\\
\hline
$0$    &$\frac{1}{4} \sqrt{\frac{5}{\pi}} (3\cos^2\theta-1)$   &$\sqrt{\frac{15}{8\pi}} \sin^2 \theta$     &$0$\\
$1$    &$\frac{1}{2} \sqrt{\frac{15}{2 \pi }} \cos\theta \sin\theta \ee^{\ii \phi } $   &   $-\frac{1}{2} \sqrt{\frac{5}{\pi}} \sin\theta \cos\theta \ee^{\ii \phi}$ & $\frac{1}{2} \sqrt{\frac{5}{\pi}} \sin\theta \ee^{\ii \phi}$ \\
$2$    &$\frac{1}{4}  \sqrt{\frac{15}{2 \pi }} \sin^2 \theta \ee^{2 \ii \phi }$ & $\frac{1}{4} \sqrt{\frac{5}{\pi}} \( 1+\cos^2 \theta \) \ee^{2\ii \phi}$ &$-\frac{1}{2} \sqrt{\frac{5}{\pi}} \cos\theta \ee^{2\ii \phi}$
\end{tabular}
\caption{Explicit representations of $Y_2^m$, $\spE^m$ and $\spB^m$.}
\label{tab:EandB}
\end{center}
\end{table}
By forming the sum and the difference of row 2 and 3 of the Boltzmann equation, we find separate evolution equations for $Q$ and $U$:
\begin{align}
\lefteqn{
\diff{}{\tau} 
\begin{pmatrix}
\Theta \\
Q\\
\ii U
\end{pmatrix}
+ \kappadot
\begin{pmatrix}
\Theta - \frac{1}{4\pi}\int{\di \Omega' \Theta'} - \nhat \cdot \vec{v}_B\\
Q \\
\ii U
\end{pmatrix}
-
\begin{pmatrix}
D_\Theta\\
0 \\
0
\end{pmatrix}
} \nonumber \\
&=
\frac{\kappadot}{10} \sum_{m=-2}^2 \int{\di \Omega'}  
\begin{pmatrix}
\Ytm \left\{\Ytmpr \Theta' -\sqrt{\frac{3}{2}} \spE^{m\prime} Q' -\sqrt{\frac{3}{2}} \spB^{m\prime} \ii U' \right\} \\
\frac{1}{2} \spE^m \left\{-\sqrt{6} \Ytmpr \Theta' +3 \spE^{m\prime} Q' + 3\spB^{m\prime} \ii U' \right\} \\
\frac{1}{2}\spB^m \left\{-\sqrt{6} \Ytmpr \Theta' +3 \spE^{m\prime} Q' + 3\spB^{m\prime} \ii U' \right\}
\end{pmatrix}. \label{eq:QandUBoltzmann}
\end{align}
Here we used the fact that the second and third entry in $\vec{D}[h_{\mu \nu}]$ vanish. The original quantities $Q\pm \ii U$ have definite spin $s=\pm 2$ so the can be expanded directly in the $\spM{\pm2}{l}{m}$ modes of equation\eqref{eq:Mlm}. This is not true for $Q$ and $U$ alone, but as we shall see, this is not a problem.
\subsection{Reducing the system of equations}
We will decompose all quantities on the left hand side of equation~\eqref{eq:QandUBoltzmann} into scalar $(m=0)$, vector $(m=\pm 1)$ and tensor $(m=\pm 2)$ components. Each component couples only to the corresponding term in the sum over collision terms. By considering the equations for $Q^{(m)}$ and $U^{(m)}$, it is clear that $\ii U^{(m)}$ and $Q^{(m)} \spB^m/\spE^m$ satisfy the same differential equation, and since the initial condition of both $Q$ and $U$ is zero, we must always have
\begin{equation}
\ii U^{(m)} = \frac{\spB^m}{\spE^m} Q^{(m)},
\end{equation}
which also covers the special case of $\spB^0=0$. (U-type polarisation vanishes for scalar modes.) This enables us to reduce the system of equations, a fact which has been used in the past for scalar modes and for tensor modes in flat space. We find
\begin{align}
\lefteqn{
\diff{}{\tau} 
\begin{pmatrix}
\Theta^{(m)} \\
Q^{(m)}
\end{pmatrix}
+ \kappadot
\begin{pmatrix}
\Theta^{(m)} - \frac{1}{4\pi}\int{\di \Omega' \Theta^{(m)\prime}} - \nhat \cdot \vec{v}_B^{(m)} \nonumber \\
Q^{(m)}
\end{pmatrix}
-
\begin{pmatrix}
D_\Theta^{(m)}\\
0
\end{pmatrix}
} \\
&=
\frac{\kappadot}{10} \int{\di \Omega'}  
\begin{pmatrix}
\Ytm \left\{\Ytmpr \Theta' -\sqrt{\frac{3}{2}} \spE^{m\prime} Q' -\sqrt{\frac{3}{2}} \spB^{m\prime} \ii U' \right\} \\
\frac{1}{2} \spE^m \left\{-\sqrt{6} \Ytmpr \Theta' +3 \spE^{m\prime} Q' + 3\spB^{m\prime} \ii U' \right\} 
\end{pmatrix} \nonumber \\
&=
\frac{\kappadot}{10} \int{\di \Omega'}  
\begin{pmatrix}
\Ytm \left\{\Ytmpr \Theta' -\sqrt{\frac{3}{2}} \[ \spE^{m\prime} + \frac{(\spB^{m\prime})^2}{\spE^{m\prime}} \]Q' \right\} \\
-\sqrt{\frac{3}{2}} \spE^m \left\{\Ytmpr \Theta' -\sqrt{\frac{3}{2}} \[ \spE^{m\prime} + \frac{(\spB^{m\prime})^2}{\spE^{m\prime}}\]Q' \right\} 
\end{pmatrix}. \label{eq:reducedBoltzmann}
\end{align}
Equation~\eqref{eq:reducedBoltzmann} shows that a single polarisation hierarchy is always enough, and this is our main result. Physically, this comes from the axial symmetry of the Thomson scattering term combined with the fact that the metric perturbations do not source polarisation directly.
\subsection{Change of variables}\label{subsec:changeofvariables}
If we do the change of variables
\begin{align}
\Theta^{(m)} &\equiv f_m(\theta) \ee^{\ii m \phi} F^{(m)}, \\
Q^{(m)} &\equiv g_m(\theta) \ee^{\ii m \phi} G^{(m)},
\end{align}
where $f_m$ and $g_m$ are two arbitrary functions to be specified later, $F^{(m)}$ and $G^{(m)}$ can be expanded in ordinary Legendre polynomials since they no longer depend on $\phi$. The form of the function $f_m$ is constrained by the requirement that the $\theta$-dependence of the following three terms can be written as a finite sum of Legendre polynomials:
\begin{itemize}
	\item the metric terms, $\ee^{-\ii m \phi} f_m^{-1} D_\Theta^{(m)}$,
	\item the Doppler term, $\ee^{-\ii m \phi} f_m^{-1} \nhat \cdot \vec{v}_B^{(m)}$,
	\item the factor in front of the $F^{(m)}$ scattering term, $\ee^{-\ii m \phi} f_m^{-1} Y_2^m$.
\end{itemize}
One can check that this requirement singles out an optimal $f_m$ up to a normalisation factor for all cases, and we give them in equation~\eqref{eq:substitutions}. The function $g_m$ is less constrained due to the absence of metric sources. The terms that must be representable by a finite number of Legendre polynomials are
\begin{itemize}
	\item the coefficient in front of $G^{(m)\prime}$: $ \[ \spE^{m\prime} + \frac{(\spB^{m\prime})^2}{\spE^{m\prime}}\] \ee^{\ii m \phi'} g_m(\theta')$,
	\item the factor in front of the $G^{(m)}$ scattering term, $\ee^{-\ii m \phi} g_m^{-1} \spE^m$.
\end{itemize}
When $m\neq 0$, the simplest function that satisfies both requirements is $g_m \sim \spE^m$. For $m=0$ we have two choices: $g_m \sim 1$ or $g_m \sim \sin^2 \theta$. The latter greatly simplifies the correspondence between $E_l^{(0)}$ and $G^{(0)}_l$, while simplifying the equations at the same time. However, the first choice is the one commonly employed in the literature on scalar polarisation (e.g. Ma\&Bertschinger~\cite{Ma:1995ey}), so we stick with this convention. In summary we have:
\begin{subequations}\label{eq:substitutions}
\begin{align}
\Theta^{(0)} &\equiv \frac{1}{4} F^{(0)}, &
Q^{(0)} &\equiv \frac{1}{4} G^{(0)}, \\
\Theta^{(1)} &\equiv \frac{1}{4} \ii \sin\theta \ee^{\ii\phi} F^{(1)}, &
Q^{(1)} &\equiv \frac{1}{4} \sin\theta\cos\theta \ee^{\ii\phi} G^{(1)}, \\
\Theta^{(2)} &\equiv \frac{1}{4} \sin^2\theta \ee^{2\ii\phi} F^{(2)}, &
Q^{(2)} &\equiv \frac{1}{4} (1+\cos^2\theta) \ee^{2\ii\phi} G^{(2)}.
\end{align}
\end{subequations}
The constant has been chosen such that $F^{(0)} \equiv F_\gamma^\text{M\&B}$ and $G^{(0)} \equiv G_\gamma^\text{M\&B}$ where $F_\gamma^\text{M\&B}$ and $G_\gamma^\text{M\&B}$ are the scalar temperature and polarisation perturbations of Ma\&Bertschinger. For tensor modes, our proposed substitution is equivalent to the one introduced by Polnarev~\cite{Polnarev:1985} and used (implicit, if not explicit) by subsequent workers,~\cite{Crittenden:1993ni,Kosowsky:1994cy,Seljak:1996is,Zaldarriaga:1996xe}. However, in these papers the substitution was always imposed \emph{before} reducing the hierarchy. We would like to emphasise that the reduction in the hierarchies has nothing to do with the variable substitution, but is a direct consequence of the structure of the Thomson scattering term in the Boltzmann equation for photons. 

We expand $F^{(m)}$ and $G^{(m)}$ in Legendre polynomials according to equation~\eqref{eq:legexpansion}. Explicitly, we have
\begin{align}
F^{(m)} = \sum_l (-\ii)^l (2l+1) F_l^{(m)} P_l, \\
G^{(m)} = \sum_l (-\ii)^l (2l+1) F_l^{(m)} P_l.
\end{align}

\subsection{Scalar, vector and tensor Boltzmann equations}
After the change of variables in the previous section, the Boltzmann equation~\eqref{eq:reducedBoltzmann} becomes
\begin{subequations}\label{eq:finalBoltzmann}
\begin{align}
\diff{}{\tau} 
\begin{pmatrix}
F^{(0)} \\
G^{(0)}
\end{pmatrix}
&+ \kappadot
\begin{pmatrix}
F^{(0)} - \int  \frac{\di\Omega'}{4\pi} F^{(0)\prime} - 4 \nhat \cdot \vec{v}_B^{(0)}\\
G^{(0)}
\end{pmatrix}
-
\begin{pmatrix}
4D_\Theta^{(0)}\\
0
\end{pmatrix}
=
\kappadot
\begin{pmatrix}
-4  P_2 \mathcal{P}^{(0)} \\
4  \[P_0-P_2 \] \mathcal{P}^{(0)}
\end{pmatrix},\label{eq:scalarBoltzmann}\\
\diff{}{\tau} 
\begin{pmatrix}
F^{(1)} \\
G^{(1)}
\end{pmatrix}
&+ \kappadot
\begin{pmatrix}
F^{(1)} - 4\frac{\ee^{-\ii\phi}}{\ii \sin\theta} \nhat \cdot \vec{v}_B^{(1)}\\
G^{(1)}
\end{pmatrix}
-
\begin{pmatrix}
4 \frac{\ee^{-\ii\phi}}{\ii \sin\theta}  D_\Theta^{(1)}\\
0
\end{pmatrix}
=
\kappadot
\begin{pmatrix}
2 \sqrt{6}  \ii P_1 \mathcal{P}^{(1)}   \\
-2 \sqrt{6}  \mathcal{P}^{(1)}   
\end{pmatrix},\label{eq:vectorBoltzmann}\\
\diff{}{\tau} 
\begin{pmatrix}
F^{(2)} \\
G^{(2)}
\end{pmatrix}
&+ \kappadot
\begin{pmatrix}
F^{(2)}\\
G^{(2)}
\end{pmatrix}
-
\begin{pmatrix}
\frac{4\ee^{-2\ii\phi}}{\sin^2\theta}  D_\Theta^{(2)}\\
0
\end{pmatrix}
=
\kappadot 
\begin{pmatrix}
-\sqrt{6} \mathcal{P}^{(2)} \\
\phantom{-} \sqrt{6} \mathcal{P}^{(2)}
\end{pmatrix}.\label{eq:tensorBoltzmann}
\end{align}
\end{subequations}
where $\mathcal{P}^{(m)}$ are given by
\begin{subequations}\label{eq:defPm}
\begin{align}
\mathcal{P}^{(0)} &= \frac{1}{8} \[ F^{(0)}_2 + G^{(0)}_0 + G^{(2)}_2 \], \\
\mathcal{P}^{(1)} &= -\frac{\sqrt{6}}{40}  \[F^{(1)}_1+F^{(1)}_3+2G^{(1)}_0+\frac{10}{7} G^{(1)}_2-\frac{4}{7}G^{(1)}_4\], \\
\mathcal{P}^{(2)} &=-\frac{1}{\sqrt{6}} \[\frac{1}{10} F^{(2)}_0 +\frac{1}{7}F^{(2)}_2+\frac{3}{70}F^{(2)}_4  - \frac{3}{5}G^{(2)}_0 +\frac{6}{7}G^{(2)}_2 -\frac{3}{70}G^{(2)}_4 \].
\end{align}
\end{subequations}
The Boltzmann equations~\eqref{eq:finalBoltzmann} are derived in detail in appendix~\ref{sec:scattering}. Equations~\eqref{eq:defPm} can a priori be taken as a definition of $\mathcal{P}^{(m)}$ but in appendix~\ref{sec:correspondence} we show that they are actually identical to the $P^{(m)}$ of Hu\&White. The linear combination of multipoles in $\mathcal{P}^{(2)}$ is usually denoted $\Psi$~\cite{Crittenden:1993ni,Seljak:1996is,weinberg2008cosmology}, while the linear combination in $\mathcal{P}^{(0)}$ is sometimes denoted by $\Pi$~\cite{Seljak:1996is}. However, we find the notation of Hu\&White more systematic. 

\subsection{The metric source term}
We split the perturbed part of the metric into scalar, vector and tensor perturbations using the eigentensors defined in equation~\eqref{eq:Qeigen} and~\eqref{eq:Qaux}. In the notation of Hu et. al., the perturbed part of the metric is written as
\begin{subequations} \label{eq:generalgauge}
\begin{align}
h_{00} &= -2 A^{(0)} \mathcal{Q}^{(0)}, \\
h_{0i} &= -B^{(0)} \mathcal{Q}_i^{(0)} -B^{(1)} \mathcal{Q}_i^{(1)}, \\
h_{ij} &= 2H_L^{(0)} \mathcal{Q}^{(0)} \gamma_{ij} + \sum_{m=0}^2 2 H_T^{(m)} \mathcal{Q}_{ij}^{(m)}.
\end{align}
\end{subequations}
By choosing a gauge, we can eliminate 2 out of the 4 scalar perturbations and one of the vector perturbations. In table~\ref{tab:gauge} we have defined the synchronous and conformal Newtonian gauge.
\begin{table}%
\begin{equation*}
\begin{array}{r@{=}l@{\qquad}r@{=}l|r@{=}l@{\qquad}r@{=}l}
\multicolumn{4}{c}{\text{Synchronous gauge}}	& \multicolumn{4}{c}{\text{Newtonian gauge}} \vrsarray \\
\hline
\quad H_L^{(0)} & \frac{1}{6}h &H_T^{(0)} &-\(3\eta + \frac{1}{2} h \)   \quad & \quad H_L^{(0)} & -\phi  &A^{(0)} & \psi \quad  \vrsarray \\
\quad H_T^{(1)} & h_V  &H_T^{(2)} & H\quad			&\quad B^{(1)} & V 	&H_T^{(2)} & H \quad \vrsarray
\end{array}
\end{equation*}
\caption{Metric perturbations in the synchronous gauge and conformal Newtonian gauge. Non-specified components in each gauge are zero. We have adopted Ma\&Bertschingers conventions for the scalar metric perturbations.}
\label{tab:gauge}
\end{table}

We now turn to the metric part of the Boltzmann equation~\eqref{eq:Boltzmann}. Only the $\Theta$-component of $\vec{D}[h_{\mu\nu}]$ is non-zero, but it is not trivial to derive in the general gauge defined by~\eqref{eq:generalgauge}. Our metric term differ from the one given in~\cite{Hu:1997hp,Hu:1997mn} by a few signs\footnote{Note that all the source terms given in both~\cite{Hu:1997hp} and \cite{Hu:1997mn} agree with our expression for $D_\Theta$.}, so we have included our derivation in appendix~\ref{sec:relBoltzmann}. We find
\begin{align}
D_\Theta &=-\frac{1}{2} n^i n^j \dot{h}_{ij} - n^i \dot{h}_{0i} + \frac{1}{2} n^i h_{00|i}.
\end{align}
We need the following identities for contraction of $\mathcal{Q}^{(m)}_{\ldots}$ by $n^i$'s: 
\begin{flalign*}
& \begin{array}{llll}
\mathcal{Q}^{(0)} =M_0^0, \qquad&
n^i \mathcal{Q}_i^{(0)} = M_1^0, \qquad &
n^i \mathcal{Q}^{(0)}_{|i} = -k M_1^0, \qquad &
n^i n^j \mathcal{Q}_{ij}^{(0)} = \frac{2}{3} \sqrt{1-3K/k^2} M_2^0 \vrarray \\
n^i \mathcal{Q}_i^{(1)} = M_1^1,\qquad &
\multicolumn{2}{l}{
n^i n^j \mathcal{Q}_{ij}^{(1)} = \frac{1}{\sqrt{3}} \sqrt{1-2K/k^2} M_2^1, 
}&
n^i n^j \mathcal{Q}_{ij}^{(2)} = M_2^2, \vrarray
\end{array}&
\end{flalign*}
where the normal modes $M_l^m$ have been defined in equation~\eqref{eq:Mlm}. Some of these relations follow directly from the definition of the normal modes of Hu et. al.~\cite{Hu:1997mn}, while the rest follow from using equation~\eqref{eq:Mderivative} in the definition of the auxiliary tensors~\eqref{eq:Qaux}. The metric part then splits into three parts,
\begin{align}
D_\Theta^{(0)} &= -\dot{H}_L^{(0)} M_0^0 + \[ kA^{(0)} +\dot{B}^{(0)} \] M_1^0 -\frac{2}{3} \sqrt{1-3K/k^2} \dot{H}_T^{(0)} M_2^0, \\
D_\Theta^{(1)} &= \dot{B}^{(1)} M_1^1 - \frac{1}{\sqrt{3}} \sqrt{1-2K/k^2} \dot{H}_T^{(1)} M_2^1, \\
D_\Theta^{(2)} &= -\dot{H}^{(2)} M_2^2.
\end{align}
The vector and tensor metric terms appearing in the Boltzmann equations~\eqref{eq:vectorBoltzmann} and~\eqref{eq:tensorBoltzmann} read explicitly:
\begin{align}
4\ee^{-\ii\phi} (\ii \sin\theta)^{-1} D_\Theta^{(1)} &= -2\sqrt{2} \dot{B}^{(1)}M_0^0 + 2\sqrt{2} \sqrt{1-2K/k^2} \dot{H}_T^{(1)} M_1^0, \\
4\ee^{-2\ii\phi} (\sin\theta)^{-2} D_\Theta^{(2)} &= \sqrt{6} \dot{H}^{(2)} M_0^0,
\end{align}
where we have used the three relations
\begin{align*}
\ee^{-\ii\phi} (\ii \sin\theta)^{-1} M_1^1 &= -\frac{1}{\sqrt{2}} M_0^0,\\
\ee^{-\ii\phi} (\ii \sin\theta)^{-1} M_2^1 &= - \sqrt{\frac{3}{2}} M_1^0, \\
\ee^{-2\ii\phi} (\sin\theta)^{-2} M_2^2 &=-\frac{1}{2}\sqrt{\frac{3}{2}} M_0^0.
\end{align*}

\subsection{Free-streaming}
Any arbitrary $\phi$-independent quantity $X(\tau,\vec{x},\hat{n})$ can be expanded in generalised Fourier modes and Legendre multipoles as
\begin{align}
X(\tau,\vec{x},\hat{n}) &= \frac{1}{(2\pi)^3} \int \di^3\vec{q} \sum_l (-\ii)^l (2l+1) X_l(\tau,\vec{q}) P_l(\mu) \Huphase \\
&= \frac{1}{(2\pi)^3} \int \di^3\vec{q} \sum_l (2l+1)X_l(\tau,\vec{q}) M_l^0,
\end{align}
and the free-streaming equation of such a function reads
\begin{equation}
\diff{X}{\tau} = \diffpar{X}{\tau} + n^i X_{|i} = 0.
\end{equation}
By using equation~\eqref{eq:Mderivative}, we find
\begin{equation}
\dot{X}_l = \frac{k}{2l+1} \[ l s_l X_{l-1} - (l+1) s_{l+1} X_{l+1} \],\qquad s_l\equiv \sqrt{1-K\frac{l^2-1}{k^2}}. \label{eq:freestreaming}
\end{equation}
Noting that $s_l=1$ for $K=0$, we can easily recover the flat limit. The hierarchy needs to be closed at some finite $l_\text{max}$, and for this Ma\&Bertschinger suggested to use the recurrence relation for spherical Bessel functions. The equivalent of spherical Bessel functions in non-flat space are hyperspherical Bessel functions. They satisfy the recurrence relation~\cite{Abbott:1986ct}
\begin{align}
\sqrt{\nu^2- \frac{Kl^2}{|K|}} \Phi_l^\nu(x) &= (2l-1) \cotk (x) \Phi_{l-1}^\nu(x) -\sqrt{\nu^2-\frac{K(l-1)^2}{|K|}}\Phi_{l-2}^\nu(x), \Rightarrow \label{eq:Phirecurrence} \\
\cotk (x) \Phi_l^\nu(x) &= \frac{1}{2l+1} \left\{ \sqrt{\nu^2-\frac{K(l+1)^2}{|K|}}\Phi_{l+1}^\nu(x) + \sqrt{\nu^2-\frac{Kl^2}{|K|}}\Phi_{l-1}^\nu(x) \right\},
\end{align}
and their derivatives can be expressed as
\begin{align}
\diff{}{x} \Phi_l^\nu(x) &= l \cotk (x) \Phi_l^\nu(x) - \sqrt{\nu^2-\frac{K(l+1)^2}{|K|}}\Phi_{l+1}^\nu(x), \\
&=\frac{1}{2l+1} \left\{ l \sqrt{\nu^2-\frac{Kl^2}{|K|}}\Phi_{l-1}^\nu(x) - (l+1) \sqrt{\nu^2-\frac{K(l+1)^2}{|K|}}\Phi_{l+1}^\nu(x) \right\}, \label{eq:ddxPhi}
\end{align}
where
\begin{equation}
\cotk (x) = \left\{
\begin{array}{ll}
\text{coth}(x) & K<0 \\
\frac{1}{x} & K=0 \\
\text{cot}(x) & K>0
\end{array}.
\right.
\end{equation}
Multiplying equation~\eqref{eq:ddxPhi} by $\sqrt{|K|}$ and using $|K|\nu^2=k^2+K$ valid for $m=0$, we find
\begin{equation}
\sqrt{|K|} \diff{}{x} \Phi_l^\nu(x) = \frac{k}{2l+1} \left\{l s_l \Phi_{l-1}^\nu(x) - (l+1) s_{l+1}\Phi_{l+1}^\nu(x) \right\},
\end{equation}
showing that $\Phi_l^\nu(x)$ with $x=\sqrt{|K|}\tau$ satisfies the free-streaming hierarchy, equation~\eqref{eq:freestreaming}. Using the recurrence relation~\eqref{eq:Phirecurrence} then leads to the following ansatz for closing the hierarchy:
\begin{equation}
\dot{X}_{\lmax} = k \[ s_{\lmax} X_{\lmax-1} - (\lmax+1)\frac{\sqrt{|K|}}{k} \cotk\(\sqrt{|K|} \tau\) X_{\lmax}\].
\end{equation}

\subsection{Boltzmann hierarchies}
After variable substitution and Legendre expansion, the Boltzmann equation becomes a hierarchy of multipoles:
\begin{align}
\dot{F}_l^{(m)} &= \frac{k}{2l+1} \[ l s_l F^{(m)}_{l-1} - (l+1) s_{l+1} F^{(m)}_{l+1} \] -\kappadot F_l^{(m)} + \sourceF_l^{(m)}, \\
\dot{G}_l^{(m)} &= \frac{k}{2l+1} \[ l s_l G^{(m)}_{l-1} - (l+1) s_{l+1} G^{(m)}_{l+1} \] -\kappadot G_l^{(m)} + \sourceG_l^{(m)},
\end{align}
where the $\sourceF_l^{(m)}$ source terms are
\begin{flalign*}
&\begin{array}{lll}
\sourceF_0^{(0)} = \kappadot F_0^{(0)}-4\dot{H}_L^{(0)}, \quad&
\sourceF_1^{(0)} = \frac{4}{3} (kA^{(0)} +\dot{B}^{(0)} + \kappadot \frac{\theta_b}{k} ) , \quad&
\sourceF_2^{(0)} = \frac{4}{5} \kappadot \mathcal{P}^{(0)} -\frac{8}{15} \sqrt{1-\frac{3K}{k^2}} \dot{H}_T^{(0)}, \vrarray\\
\sourceF_0^{(1)} =  -\frac{1}{\sqrt{2}} (\dot{B}^{(1)}+4 v_B^{(1)}), &
\multicolumn{2}{l}{ 
\sourceF_1^{(1)} = \frac{4}{3\sqrt{2}} \sqrt{1-2K/k^2} \dot{H}_T^{(1)} - 2\sqrt{\frac{2}{3}} \kappadot \mathcal{P}^{(1)},
} \vrarray \\
\multicolumn{3}{l}{
\sourceF_0^{(2)} = 2 \sqrt{\frac{3}{2}} \dot{H}_T^{(2)} -2\sqrt{\frac{3}{2}} \kappadot \mathcal{P}^{(2)},\vrarray
}
\end{array}&
\end{flalign*}
while the polarisation source terms $\sourceG_l^{(m)}$ are given in terms of the quantities $\mathcal{P}^{(m)}$ of equation~\eqref{eq:defPm}:
\begin{flalign*}
& \sourceG_0^{(0)} = 4 \kappadot \mathcal{P}^{(0)}, \qquad 
\sourceG_2^{(0)} = \frac{4}{5} \kappadot \mathcal{P}^{(0)}, \qquad 
\sourceG_0^{(1)} = 4\sqrt{\frac{3}{2}} \kappadot \mathcal{P}^{(1)}, \qquad
\sourceG_0^{(2)} = 2\sqrt{\frac{3}{2}} \kappadot \mathcal{P}^{(2)}. &
\end{flalign*}
In the synchronous and conformal gauge, $\sourceF_l^{(m)}$ reduce to
\paragraph{Newtonian gauge}
\begin{flalign*}
&\begin{array}{lll}
\sourceF_0^{(0)} = 4\dot{\phi}+\kappadot F_0^{(0)},\qquad&
\sourceF_1^{(0)} = \frac{4}{3} k \psi + \kappadot \frac{4}{3k} \theta_b , \qquad &
\sourceF_2^{(0)} = \frac{4}{5} \kappadot \mathcal{P}^{(0)} , \vrarray \\
\sourceF_0^{(1)} =  -2\sqrt{2} \dot{V}-2\sqrt{2}\kappadot v_B^{(1)}, \qquad&
\sourceF_1^{(1)} = - 2\sqrt{\frac{2}{3}} \kappadot \mathcal{P}^{(1)}, \vrarray \\
\multicolumn{3}{l}{
\sourceF_0^{(2)} = 2 \sqrt{\frac{3}{2}} \dot{H} -2\sqrt{\frac{3}{2}} \kappadot \mathcal{P}^{(2)}.
} \vrarray
\end{array}&
\end{flalign*}
\paragraph{Synchronous gauge}
\begin{flalign*}
&\begin{array}{lll}
\sourceF_0^{(0)} = -\frac{2}{3} \dot{h}+\kappadot F_0^{(0)},\qquad &
\sourceF_1^{(0)} = \kappadot \frac{4}{3k} \theta_b, \qquad &
\sourceF_2^{(0)} = \frac{4}{15} \sqrt{1-3K/k^2} (6\dot{\eta}+\dot{h})+\frac{4}{5} \kappadot \mathcal{P}^{(0)}, \vrarray \\
\sourceF_0^{(1)} =  -2\sqrt{2}\kappadot v_B^{(1)},& 
\multicolumn{2}{l}{ 
\sourceF_1^{(1)} = \frac{4}{3\sqrt{2}} \sqrt{1-2K/k^2} \dot{h}_V - 2\sqrt{\frac{2}{3}} \kappadot \mathcal{P}^{(1)}, 
} \vrarray \\
\multicolumn{3}{l}{
\sourceF_0^{(2)} = 2 \sqrt{\frac{3}{2}} \dot{H} -2\sqrt{\frac{3}{2}} \kappadot \mathcal{P}^{(2)}.
} \vrarray
\end{array}&
\end{flalign*}

\subsection{The line of sight integrals}
The line of sight integrals are most easily derived in the original variables of Hu et. al.~\cite{Hu:1997mn}. We will just quote their general results here:
\begin{align}
\frac{\Theta_l^{(m)}}{2l+1} &=\int_0^{\tau_0} \di \tau \ee^{-\kappa} \sum_j S_j^{(m)} \phi_l^{(jm)}, \\
\frac{E_l^{(m)}}{2l+1} &=\int_0^{\tau_0} \di \tau \kappadot  \ee^{-\kappa} \( -\sqrt{6} \mathcal{P}^{(m)} \) \epsilon_l^{(m)}, \\
\frac{B_l^{(m)}}{2l+1} &=\int_0^{\tau_0} \di \tau \kappadot  \ee^{-\kappa} \( -\sqrt{6} \mathcal{P}^{(m)} \) \beta_l^{(m)},
\end{align}
where $\phi_l$, $\epsilon_l$ and $\beta_l$ are the radial functions of Hu et. al.~\cite{Hu:1997mn}. In a flat Universe, they are given in terms of spherical Bessel functions, and in the non-flat case they are given in terms of hyperspherical Bessel functions. The source for polarisation, $\mathcal{P}^{(m)}$, are given by equation~\eqref{eq:defPm}, while the $S_j^{(m)}$ are given by
\begin{equation*}
S_j^{(0)} = \frac{2j+1}{4} \sourceF_j^{(0)}, \qquad 
S_1^{(1)} = -\frac{1}{\sqrt{8}} \sourceF_0^{(1)}, \qquad S_2^{(1)} = -\sqrt{\frac{3}{8}} \sourceF_1^{(1)}, \qquad
S_2^{(2)} = -\frac{1}{\sqrt{6}} \sourceF_0^{(2)}.
\end{equation*}

%

\subsection{Correspondence formulae}
We only need the $E_2^{(m)}$ multipole to get the line-of-sight integrals, but in appendix~\ref{sec:correspondence} we have derived the general formulae for recovering the $E_l^{(m)}$ and $B_l^{(m)}$ multipoles from our polarisation multipoles $G_l^{(m)}$. The computations are rather lengthy, but the final relations are:
\begin{subequations}
\begin{align}
\Theta_l^{(0)} &= \frac{(2l+1)}{4} F_l^{(0)}, \\
E_l^{(0)} &= \sqrt{\frac{(l-2)!}{(l+2)!}} \frac{2l+1}{4} \[-l(l-1) G^{(0)}_l+ \sum_{k=0}^{l-2} \ii^{l-k} \(1+(-1)^{l+k}\) (2k+1) G^{(0)}_k  \], \\
\Theta_l^{(1)} &= -\frac{1}{4} \sqrt{\frac{(l+1)!}{(l-1)!}} \(F_{l-1}^{(1)} + F_{l+1}^{(1)} \), \\
E_l^{(1)} &= \frac{1}{2}\sqrt{\frac{2l+1}{5}} \[ (2l-3) \gamma_{l-2}^l G^{(1)}_{l-2} - (2l+1) \gamma_{l}^{l} G^{(1)}_l + (2l+5) \gamma_{l+2}^{l} G^{(1)}_{l+2} \], \\
B_l^{(1)} &= \frac{1}{2}\sqrt{\frac{2l+1}{5}} \[ -(2l-1)\gamma_{l-1}^l G^{(1)}_{l-1} + (2l+3)\gamma_{l+1}^l G^{(1)}_{l+1} \], \\
\Theta^{(2)}_l &= -\frac{1}{4} \sqrt{\frac{(l+2)!}{(l-2)!}} \[\frac{1}{2l-1} F^{(2)}_{l-2} + \frac{2(2l+1)}{(2l-1)(2l+3)} F^{(2)}_l + \frac{1}{2l+3} F^{(2)}_{l+2} \], \\
E_l^{(2)} &= \sqrt{\frac{2l+1}{5}} \[-(2l-3) G^{(2)}_{l-2} \alpha_{l-2}^l + (2l+1)G^{(2)}_l \alpha_l^l -(2l+5) G^{(2)}_{l+2} \alpha_{l+2}^l \], \\
B_l^{(2)} &= \sqrt{\frac{2l+1}{5}} \[ (2l-1)G^{(2)}_{l-1}\alpha_{l-1}^l - (2l+3) G^{(2)}_{l+1} \alpha_{l+1}^l \].
\end{align}
\end{subequations}
Closed form expressions for $\gamma_l^j$ and $\alpha_l^j$ are given in equation~\eqref{eq:gammalj} and~\eqref{eq:alphalj} respectively.

\section{Conclusion}
In this paper we have showed that calculating the CMB polarisation by evolving the $E$- and $B$-mode as it is usually done is not optimal. Instead, evolving a single quantity, $G^{(m)}$, is enough, and the multipoles of $E^{(m)}$ and $B^{(m)}$ can then be recovered from the multipoles of $G^{(m)}$. This was previously known only for scalar perturbations and tensor perturbations in flat space. In addition to the obvious computational advantage of having one less hierarchy to evolve in time, the free-streaming solution is also simplified in our approach since all perturbations are expanded in ordinary Legendre polynomials. These equations will soon be implemented in the public code \CLASS{}\footnote{Available at \url{http://class-code.net}}.

\acknowledgments{This project is supported by the Swiss National Foundation. We wish to thank Simon Prunet for very stimulating discussions.}

\appendix

\section{Scattering terms}\label{sec:scattering}
In this appendix we show how to calculate the scattering terms in equations~\eqref{eq:finalBoltzmann}.
\subsection{Scalar perturbations}
After the substitution 
\begin{align}
\Theta^{(0)} &\equiv \frac{1}{4} F^{(0)}, \\
Q^{(0)} &\equiv \frac{1}{4} G^{(0)},
\end{align}
equation~\eqref{eq:reducedBoltzmann} for $m=0$ reads
\begin{align}
\lefteqn{
\diff{}{\tau} 
\begin{pmatrix}
F^{(0)} \\
G^{(0)}
\end{pmatrix}
+ \kappadot
\begin{pmatrix}
F^{(0)} - \frac{1}{4\pi}\int{\di \Omega'} F^{(0)\prime} - 4 \nhat \cdot \vec{v}_B^{(0)}\\
G^{(0)}
\end{pmatrix}
-
\begin{pmatrix}
4D_\Theta^{(0)}\\
0
\end{pmatrix}
} \nonumber\\
&=
\frac{\kappadot}{10} \int{\di \Omega'}  
\begin{pmatrix}
\Ytn \left\{\Ytnpr F^{(0)\prime} -\sqrt{\frac{3}{2}} \[ \spE^{0\prime} + \frac{(\spB^{0\prime})^2}{\spE^{0\prime}} \]G^{(0)\prime} \right\} \\
-\sqrt{\frac{3}{2}} \spE^0 \left\{\Ytnpr F^{(0)\prime} -\sqrt{\frac{3}{2}} \[ \spE^{0\prime} + \frac{(\spB^{0\prime})^2}{\spE^{0\prime}}\] G^{(0)\prime} \right\} 
\end{pmatrix}\nonumber\\
&=
\frac{\kappadot}{10} \int{\di \Omega'}  
\begin{pmatrix}
\frac{1}{2}\sqrt{\frac{5}{\pi}} P_2 \left\{\frac{1}{2}\sqrt{\frac{5}{\pi}} P_2' F^{(0)\prime} -\sqrt{\frac{3}{2}} \sqrt{\frac{5}{6\pi}}\[P_0'-P_2' \]G^{(0)\prime} \right\} \\
-\sqrt{\frac{3}{2}} \sqrt{\frac{5}{6\pi}}\[P_0-P_2 \] \left\{\frac{1}{2}\sqrt{\frac{5}{\pi}} P_2' F^{(0)\prime} -\sqrt{\frac{3}{2}} \sqrt{\frac{5}{6\pi}}\[P_0'-P_2' \]G^{(0)\prime} \right\}
\end{pmatrix} \nonumber\\
&=
\frac{\kappadot}{10} \int_{-1}^{1}{\di \mu'}  
\begin{pmatrix}
5 P_2 \left\{\frac{1}{2} P_2' F^{(0)\prime} -\frac{1}{2} \[P_0'-P_2' \]G^{(0)\prime} \right\} \\
-5\[P_0-P_2 \] \left\{\frac{1}{2} P_2' F^{(0)\prime} -\frac{1}{2} \[P_0'-P_2' \]G^{(0)\prime} \right\}
\end{pmatrix} \nonumber\\
&=
\frac{\kappadot}{10} \int_{-1}^{1}{\di \mu'}  
\begin{pmatrix}
5 P_2 \left\{\frac{1}{2} P_2' F^{(0)\prime} -\frac{1}{2} \[P_0'-P_2' \]G^{(0)\prime} \right\} \\
-5\[P_0-P_2 \] \left\{\frac{1}{2} P_2' F^{(0)\prime} -\frac{1}{2} \[P_0'-P_2' \]G^{(0)\prime} \right\}
\end{pmatrix} \nonumber\\
&=
\frac{\kappadot}{10}  
\begin{pmatrix}
-5 P_2 \left\{F^{(0)}_2 + G^{(0)}_0 + G^{(0)}_2 \right\} \\
5\[P_0-P_2 \] \left\{F^{(0)}_2 + G^{(0)}_0 + G^{(0)}_2 \right\}
\end{pmatrix}\nonumber\\
&=
\kappadot
\begin{pmatrix}
-4  P_2 \mathcal{P}^{(0)} \\
4  \[P_0-P_2 \] \mathcal{P}^{(0)}
\end{pmatrix}.
\end{align}

\subsection{Vector perturbations}
For the vector perturbations, we do the following change of variables:
\begin{align}
\Theta^{(1)} &\equiv \frac{1}{4} \ii \sin\theta \ee^{\ii\phi} F^{(1)} = \frac{1}{2} \ii \sqrt{\frac{2\pi}{15}} \frac{1}{\cos \theta} Y_2^1 F^{(1)}, \\
Q^{(1)} &\equiv \frac{1}{4} \sin\theta\cos\theta \ee^{\ii\phi} G^{(1)} = -\frac{1}{2} \sqrt{\frac{\pi}{5}} \spE^1 G^{(1)}. 
\end{align}
In these variables, the Boltzmann equation~\eqref{eq:reducedBoltzmann} takes the form
\begin{align}
\lefteqn{
\diff{}{\tau} 
\begin{pmatrix}
F^{(1)} \\
G^{(1)}
\end{pmatrix}
+ \kappadot
\begin{pmatrix}
F^{(1)} - 4\frac{\ee^{-\ii\phi}}{\ii \sin\theta} \nhat \cdot \vec{v}_B^{(1)}\\
G^{(1)}
\end{pmatrix}
-
\begin{pmatrix}
4 \frac{\ee^{-\ii\phi}}{\ii \sin\theta}  D_\Theta^{(1)}\\
0
\end{pmatrix}
} \nonumber\\
&=
\frac{\kappadot}{10} \int{\di \Omega'}  
\begin{pmatrix}
\phantom{-} \frac{1}{2} \sqrt{\frac{15}{2\pi}} (-\ii) \cos\theta \left\{\Ytepr \ii \sin\theta' \ee^{\ii\phi'} F^{(1)\prime} -\sqrt{\frac{3}{2}} \[ \spE^{1\prime} + \frac{(\spB^{1\prime})^2}{\spE^{1\prime}} \] \sin\theta\cos\theta \ee^{\ii\phi'} G^{(1)\prime} \right\} \\
\frac{1}{2}\sqrt{\frac{15}{2\pi}} \left\{\Ytepr \ii \sin\theta' \ee^{\ii\phi'} F^{(1)\prime} -\sqrt{\frac{3}{2}} \[ \spE^{1\prime} + \frac{(\spB^{1\prime})^2}{\spE^{1\prime}} \] \sin\theta\cos\theta \ee^{\ii\phi'} G^{(1)\prime} \right\} 
\end{pmatrix}\nonumber\\
&=
\frac{\kappadot}{10} \int{\di \Omega'}  
\begin{pmatrix}
\frac{1}{2} \sqrt{\frac{15}{2\pi}} (-\ii) \mu \left\{ \(  \frac{1}{2}\sqrt{\frac{15}{2\pi}} \mu' (1-\mu^{\prime 2})  \) \ii F^{(1)\prime} +  \(\frac{1}{2} \sqrt{\frac{15}{2 \pi }} (1-\mu^{\prime 4}) \) G^{(1)\prime} \right\} \\
\frac{1}{2}\sqrt{\frac{15}{2\pi}} \left\{ \(  \frac{1}{2}\sqrt{\frac{15}{2\pi}}\mu' (1-\mu^{\prime 2})  \) \ii F^{(1)\prime} +  \(\frac{1}{2} \sqrt{\frac{15}{2 \pi }} (1-\mu^{\prime 4}) \) G^{(1)\prime} \right\}
\end{pmatrix} \nonumber\\
&=
\frac{\kappadot}{10} \int{\di \Omega'}  
\begin{pmatrix}
\frac{1}{4} \frac{15}{2\pi} (-\ii)  \mu \left\{ \( \mu' (1-\mu^{\prime 2})  \)\ii F^{(1)\prime} +  \( 1-\mu^{\prime 4} \) G^{(1)\prime} \right\} \\
\frac{1}{4}\frac{15}{2\pi} \left\{ \( \mu' (1-\mu^{\prime 2})  \)\ii F^{(1)\prime} +  \(1-\mu^{\prime 4} \) G^{(1)\prime} \right\}
\end{pmatrix} \nonumber\\
&=
\frac{\kappadot}{10} \int{\di \Omega'}  
\begin{pmatrix}
 \frac{1}{4} \frac{15}{2\pi} (-\ii) P_1 \left\{ \( \frac{2}{5}P_1' -\frac{2}{5} P_3' \)\ii F^{(1)\prime} +  \( \frac{4}{5}P_0' - \frac{4}{7}P_2' -\frac{8}{35} P_4' \) G^{(1)\prime} \right\} \\
\frac{1}{4}\frac{15}{2\pi} \left\{ \( \frac{2}{5}P_1' -\frac{2}{5} P_3' \)\ii F^{(1)\prime} +  \( \frac{4}{5}P_0' - \frac{4}{7}P_2' -\frac{8}{35} P_4' \) G^{(1)\prime} \right\}
\end{pmatrix} \nonumber\\
&=
\frac{\kappadot}{10}  
\begin{pmatrix}
 \frac{15}{2} (-\ii) P_1 \left\{  \frac{2}{5}F^{(1)}_1 + \frac{2}{5}F^{(1)}_3 +  \frac{4}{5}G^{(1)}_0 + \frac{4}{7}G^{(1)}_2 -\frac{8}{35} G^{(1)}_4 \right\} \\
\frac{15}{2} \left\{  \frac{2}{5}F^{(1)}_1 + \frac{2}{5}F^{(1)}_3 +  \frac{4}{5}G^{(1)}_0 + \frac{4}{7}G^{(1)}_2 -\frac{8}{35} G^{(1)}_4 \right\}
\end{pmatrix}\nonumber\\
&=
\frac{\kappadot}{10}  
\begin{pmatrix}
3(-\ii) P_1 \left\{  F^{(1)}_1 +  F^{(1)}_3 +  2 G^{(1)}_0 + \frac{10}{7}G^{(1)}_2 -\frac{4}{7} G^{(1)}_4 \right\} \\
 3\left\{  F^{(1)}_1 +  F^{(1)}_3 +  2 G^{(1)}_0 + \frac{10}{7}G^{(1)}_2 -\frac{4}{7} G^{(1)}_4 \right\}
\end{pmatrix}\nonumber \\
&=
\kappadot
\begin{pmatrix}
2 \sqrt{6}  \ii P_1 \mathcal{P}^{(1)}   \\
-2 \sqrt{6}  \mathcal{P}^{(1)}   
\end{pmatrix}.
\end{align}
\subsection{Tensor perturbations}
For the tensor perturbations, we do the change of variables
\begin{align}
\Theta^{(2)} &\equiv \frac{1}{4} \sin^2\theta \ee^{2\ii\phi} F^{(2)} = \sqrt{\frac{2\pi}{15}} Y_2^2 F^{(2)}, \\
Q^{(2)} &\equiv \frac{1}{4} (1+\cos^2\theta) \ee^{2\ii\phi} G^{(2)} = \sqrt{\frac{3}{2}} \sqrt{\frac{2\pi}{15}} \spE^2 G^{(2)}.
\end{align}
After this substitution, the Boltzmann equation~\eqref{eq:reducedBoltzmann} reads
\begin{align}
\lefteqn{
\diff{}{\tau} 
\begin{pmatrix}
F^{(2)} \\
G^{(2)}
\end{pmatrix}
+ \kappadot
\begin{pmatrix}
F^{(2)}\\
G^{(2)}
\end{pmatrix}
-
\begin{pmatrix}
\frac{4\ee^{-2\ii\phi}}{\sin^2\theta}  D_\Theta^{(2)}\\
0
\end{pmatrix}
} \nonumber \\
&=
\frac{\kappadot}{10} \int{\di \Omega'}  
\begin{pmatrix}
\phantom{-} \frac{1}{4}\sqrt{\frac{15}{2\pi}} \left\{\Yttpr \Theta' -\sqrt{\frac{3}{2}} \[ \spE^{2\prime} + \frac{(\spB^{2\prime})^2}{\spE^{2\prime}} \]Q' \right\} \\
-\frac{1}{4}\sqrt{\frac{15}{2\pi}} \left\{\Yttpr \Theta' -\sqrt{\frac{3}{2}} \[ \spE^{2\prime} + \frac{(\spB^{2\prime})^2}{\spE^{2\prime}}\]Q' \right\} 
\end{pmatrix}\nonumber\\
&=
\frac{\kappadot}{10} \int{\di \Omega'}  
\begin{pmatrix}
\phantom{-} \frac{1}{16}\frac{15}{2\pi} \left\{\(1-\mu^{\prime 2} \)^2 F^{(2)\prime} - \( 1 + 6 \mu^{\prime 2} + \mu^{\prime 4} \)G^{(2)\prime} \right\} \\
- \frac{1}{16}\frac{15}{2\pi} \left\{\(1-\mu^{\prime 2} \)^2 F^{(2)\prime} - \( 1 + 6 \mu^{\prime 2} + \mu^{\prime 4} \)G^{(2)\prime} \right\}  
\end{pmatrix}\nonumber\\
&=
\frac{\kappadot}{10} \int_{-1}^1 {\di \mu'}  
\begin{pmatrix}
\phantom{-} \frac{1}{2} \frac{15}{8} \left\{\( \frac{8}{15} P_0' -\frac{16}{21}P_2'+\frac{8}{35}P_4' \) F^{(2)\prime} - \( \frac{16}{5}P_0' +\frac{32}{7}P_2' +\frac{8}{35}P_4' \)G^{(2)\prime} \right\} \\
- \frac{1}{2}\frac{15}{8} \left\{\( \frac{8}{15} P_0' -\frac{16}{21}P_2'+\frac{8}{35}P_4' \) F^{(2)\prime} - \( \frac{16}{5}P_0' +\frac{32}{7}P_2' +\frac{8}{35}P_4' \)G^{(2)\prime} \right\}  
\end{pmatrix}\nonumber\\
&=
\frac{\kappadot}{10} \int_{-1}^1 {\di \mu'} 
\begin{pmatrix}
\phantom{-} \frac{1}{2} \left\{\(  P_0' -\frac{10}{7}P_2'+\frac{3}{7}P_4' \) F^{(2)\prime} - \( 6P_0' +\frac{60}{7}P_2' +\frac{3}{7}P_4' \)G^{(2)\prime} \right\} \\
- \frac{1}{2} \left\{\(  P_0' -\frac{10}{7}P_2'+\frac{3}{7}P_4' \) F^{(2)\prime} - \( 6P_0' +\frac{60}{7}P_2' +\frac{3}{7}P_4' \)G^{(2)\prime} \right\}  
\end{pmatrix}\nonumber\\
&=
\frac{\kappadot}{10} 
\begin{pmatrix}
\phantom{-}  \left\{  F^{(2)}_0 +\frac{10}{7}F^{(2)}_2+\frac{3}{7}F^{(2)}_4  - 6G^{(2)}_0 +\frac{60}{7}G^{(2)}_2 -\frac{3}{7}G^{(2)}_4 \right\} \\
- \left\{  F^{(2)}_0 +\frac{10}{7}F^{(2)}_2+\frac{3}{7}F^{(2)}_4  - 6G^{(2)}_0 +\frac{60}{7}G^{(2)}_2 -\frac{3}{7}G^{(2)}_4 \right\}  
\end{pmatrix}\nonumber\\
&=
\kappadot 
\begin{pmatrix}
-\sqrt{6} \mathcal{P}^{(2)} \\
\phantom{-} \sqrt{6} \mathcal{P}^{(2)}
\end{pmatrix}.
\end{align}

\section{Correspondence between expansions}\label{sec:correspondence}
Here we derive the correspondence between the ordinary Legendre expansion coefficients $F_l^{(m)}$ and $G_l^{(m)}$ and the coefficients $\Theta_l^{(m)}$, $E_l^{(m)}$ and $B_l^{(m)}$ in the spin-weighted spherical harmonics expansions of Hu et. al.
\subsection{Scalar modes}
The scalar temperature expansions can be directly compared. Hu et. al.'s expansion is
\begin{equation}
\Theta^{(0)} = \sum_l (-\ii)^l \Theta^{(0)}_l \sqrt{\frac{4\pi}{2l+1}} Y_l^0 = \sum_l (-\ii)^l \Theta^{(0)}_l P_l,
\end{equation}
while ours is
\begin{equation}
\Theta^{(0)} = \frac{1}{4} F^{(0)} = \sum_l (-\ii)^l (2l+1) \frac{1}{4} F^{(0)}_l P_l,
\end{equation}
so we find
\begin{equation}
\Theta_l^{(0)} = \frac{(2l+1)}{4} F_l^{(0)}.
\end{equation}

The $E_l^{0}$ relation is more complicated. (Remember that $B_l^{(0)}$ is zero.) We use the following explicit formula for $\spY{\pm2}{l}{0}$:
\begin{equation}
\spY{\pm2}{l}{0} = \sqrt{\frac{(l-2)!}{(l+2)!}} \sqrt{\frac{2l+1}{4\pi}} P_l^2,
\end{equation}
which is easily derived from equation (3) in~\cite{Hu:1997hp}. Hu et. al.'s expansion can then be written as
\begin{equation}
Q^{(0)} = Q^{(0)}\pm \ii U^{(0)} =\sum_l (-\ii)^l \sqrt{\frac{(l-2)!}{(l+2)!}} E_l^{(0)} P_l^2, 
\end{equation}
which leads to
\begin{align}
E_l^{(0)} &= (-\ii)^{-l} \sqrt{\frac{(l-2)!}{(l+2)!}} \frac{2l+1}{2} \int_{-1}^{1} \di \mu Q^{(0)} P_l^2, \nonumber \\
&= (-\ii)^{-l} \sqrt{\frac{(l-2)!}{(l+2)!}} \frac{2l+1}{2} \sum_k (-\ii)^k \frac{2k+1}{4} G^{(0)}_k \int_{-1}^1 \di \mu P_k P_l^2. \label{eq:El0sum}
\end{align}
We could not find the general formula for the needed integral\footnote{However, had we instead chosen $Q^{(0)} \sim \sin^2\theta G^{(0)}$ as discussed in section~\ref{subsec:changeofvariables}, this integral would have been given directly by Gaunt's formula.} anywhere in the literature, so we have included its derivation here. The first step is to use the definition of the associated Legendre polynomial, equation~\eqref{eq:Plmdef}:
\begin{align}
\lefteqn{\int_{-1}^{1} \di \mu P_k P_l^2 = \int_{-1}^{1} \di \mu (1-\mu^2) P_k P_l^{\prime\prime} }\nonumber\\
&=\int_{-1}^{1} \di \mu \[-\frac{k(k-1) P_{k-2}}{(2k-1) (2k+1)} + \frac{2 (k^2+k-1)P_k}{(2k-1) (2k+3)}-\frac{(k+1) (k+2)P_{k+2}}{(2k+1) (2k+3)} \] P_l^{\prime\prime} \label{eq:intPlPl2}.
\end{align}
The second derivative of a Legendre polynomial can be recast into a sum over Legendre polynomials. To derive this sum, we start from the basic relation
\begin{equation}
(2n+1)P_n = P_{n+1}'-P_{n-1}',
\end{equation}
which can be iterated to give
\begin{equation}
P_n'=\sum_{j=0}^{n-1} \(1-(-1)^{j+n} \) \frac{2j+1}{2} P_j.
\end{equation}
The second derivative can then be expressed as
\begin{align*}
P_n^{\prime\prime} &= \sum_{j=0}^{n-1} \(1-(-1)^{j+n} \) \frac{2j+1}{2} \sum_{k=0}^{j-1} \(1-(-1)^{k+j} \) \frac{2k+1}{2} P_k \\
&=\sum_{j=0}^{n-1}  \sum_{k=0}^{j-1} \(1-(-1)^{k+j} \) \(1-(-1)^{j+n} \) \frac{2j+1}{2} \frac{2k+1}{2} P_k \\
&=\sum_{k=0}^{n-2}  \sum_{j=k+1}^{n-1} \(1-(-1)^{k+j} \) \(1-(-1)^{j+n} \) \frac{2j+1}{2} \frac{2k+1}{2} P_k \\
&=\sum_{k=0}^{n-2} \(1+(-1)^{k+n} \) \frac{2k+1}{4} (n-k) (n+k+1) P_k.
\end{align*}
This leads to
\begin{align}
\int_{-1}^{1} \di \mu P_k P_l^{\prime\prime} &=\sum_{j=0}^{l-2} \(1+(-1)^{j+l} \) \frac{2j+1}{4} (l-j) (l+j+1) \frac{2}{2k+1} \delta_{jl} \nonumber \\
&=
\left\{
\begin{array}{ll}
	\frac{1}{2} \(1+(-1)^{k+n} \) (l-k) (l+k+1), & \quad k\leq l-2, \nonumber \\
	0,                                           & \quad k>l-2.
\end{array}
\right.\label{eq:intPkPlprpr}
\end{align}
We now insert equation~\eqref{eq:intPkPlprpr} that we just derived in equation~\eqref{eq:intPlPl2}. We find
\begin{align}
\int_{-1}^{1} \di \mu P_k P_l^2 = \frac{1+(-1)^{l+k}}{2}  &\left\{ -\frac{k(k-1) (l-(k-2)) (l+(k-2)+1) \I_{k\leq l}}{(2k-1) (2k+1)} + \right. \nonumber\\
&+\frac{2 (k^2+k-1)(l-k) (l+k+1)\I_{k\leq l-2}}{(2k-1) (2k+3)}- \nonumber\\
&\left.-\frac{(k+1) (k+2) (l-(k+2)) (l+(k+2)+1) \I_{k\leq l-4}}{(2k+1) (2k+3)} \right\} \nonumber\\
&=\left\{
\begin{array}{ll}
4 & \text{if }k \leq l-2 \text{ and } k+l\text{ is even,}\\
-\frac{2 l(l-1)}{2l+1} &\text{for }k=l, \\ 
$0$&\text{otherwise.}
\end{array}.
\right.
\end{align}
We can now evaluate the integrals in equation~\eqref{eq:El0sum}. We find that the correspondence between $E_l^{(0)}$ and $G_l^{(0)}$ reads:
\begin{align}
E_l^{(0)} &= (-\ii)^{-l} \sqrt{\frac{(l-2)!}{(l+2)!}} \frac{2l+1}{4} \[-(-\ii)^l l(l-1) G^{(0)}_l+ \sum_{k=0}^{l-2} (-\ii)^k \(1+(-1)^{l+k}\) (2k+1) G^{(0)}_k  \], \nonumber \\
&= \sqrt{\frac{(l-2)!}{(l+2)!}} \frac{2l+1}{4} \[-l(l-1) G^{(0)}_l+ \sum_{k=0}^{l-2} \ii^{l-k} \(1+(-1)^{l+k}\) (2k+1) G^{(0)}_k  \].
\end{align}
We can now calculate the quantity $P^{(0)}$ of Hu\&White starting from their definition
\begin{equation}\label{eq:HuPm}
P^{(m)} = \frac{1}{10} \[ \Theta_2^{(m)} -\sqrt{6} E_2^{(m)} \].
\end{equation}
We have $\Theta_2^{(0)} = \frac{5}{4} F_2^{(0)}$ and
\begin{equation*}
\sqrt{6}E_l^{(0)} = \frac{\sqrt{6}}{\sqrt{4!}} \frac{5}{4} \[-2 G_2^{(0)} -2 G_0^{(0)} \] = -\frac{5}{4} \[G_0^{(0)} + G_2^{(0)} \],
\end{equation*}
so we find
\begin{equation}
P^{(0)} = \frac{1}{8} \[ F_0^{(0)} +  G_0^{(0)} + G_2^{(0)} \],
\end{equation}
in accordance with our definition of $\mathcal{P}^{(0)}$, equation~\eqref{eq:defPm}. 

In the notation of Seljak\&Zaldarriaga~\cite{Seljak:1996is}, we have $P^{(0)}=\frac{1}{2} \Pi$ where we have used the correspondence $F_l^{(0)}=4\Delta_{T,l}^{(S)}$ and $G_l^{(0)}=4\Delta_{P,l}^{(S)}$ between our multipoles and their multipoles.

\subsection{Vector modes}
For connecting $F^{(1)}$ and $\Theta^{(1)}$, we use the formula
\begin{equation}
\sqrt{1-\mu^2} P_l = \frac{1}{2l+1} \[ P_{l+1}^1-P_{l-1}^1 \].
\end{equation}
Note that this equation depends on the convention for the Condon-Shortley phase. We rewrite our expansion of $F^{(1)}$ in the following way:
\begin{align*}
\Theta^{(1)} &= \frac{1}{4} \ii \sin\theta \ee^{\ii\phi} \sum_l (-\ii)^l (2l+1) F^{(1)}_l P_l, \\
&= \sum_l (-\ii)^{l-1} \frac{1}{4} F^{(1)}_l \ee^{\ii\phi} \(P_{l+1}^1-P_{l-1}^1 \), \\
&= \sum_l \frac{1}{4} \( (-\ii)^{l-2}  F^{(1)}_{l-1} - (-\ii)^{l}  F^{(1)}_{l+1} \) P_l^1 \ee^{\ii\phi},  \\
&= \sum_l (-\ii)^l \frac{1}{4} \sqrt{\frac{4\pi}{2l+1}} \sqrt{\frac{(l+1)!}{(l-1)!}} \(-F^{(1)}_{l-1} - F^{(1)}_{l+1} \) Y_l^1.
\end{align*}
Comparing this to the expansion of Hu\&White yields
\begin{equation}
\Theta_l^{(1)} = -\frac{1}{4} \sqrt{\frac{(l+1)!}{(l-1)!}} \(F_{l-1}^{(1)} + F_{l+1}^{(1)} \).
\end{equation}
We now turn to the $E^{(1)}_l$ and $B^{(1)}_l$ relations. We have
\begin{align*}
Q^{(1)}+\ii U^{(1)} &= Q^{(1)} + \frac{\spB^1}{\spE^1} Q^{(1)}, \\
&= \frac{1}{4} \(1-\frac{1}{\cos\theta} \) \sin\theta \cos\theta \ee^{\ii\phi} G^{(1)}, \\
&= -\sqrt{\frac{\pi}{5}} \spY{2}{2}{1} \sum_l (-\ii)^l (2l+1) G^{(1)}_l P_l.
\end{align*}
The product $P_l \spY{2}{2}{1}$ can be expanded in $s=2, m=1$ spin-weighted spherical harmonics:
\begin{equation}
P_l \spY{2}{2}{1} = \sum_j \gamma_l^j \spY{2}{j}{1},
\end{equation}
with expansion coefficients given by
\begin{align}
\gamma_l^j &= \sqrt{\frac{4\pi}{2l+1}} \int \di \Omega \spY{2}{2}{1} \spY{0}{l}{0} (\spY{2}{j}{1})^* = -\sqrt{\frac{4\pi}{2l+1}} \int \di \Omega \spY{2}{2}{1} \spY{0}{l}{0} \spY{-2}{j}{-1} \nonumber \\
&= -\sqrt{5(2j+1)} \Wthreej{2}{1}{l}{0}{j}{-1} \Wthreej{2}{-2}{l}{0}{j}{2}.
\end{align}
The selection rules for the Wigner 3-j symbols tells us that $\gamma_l^j$ is non-zero if and only if $j\geq 2$ and $|l-2|\leq j \leq l+2$. In this range, $\gamma_l^j$ can be written as\footnote{We have used equation~\eqref{eq:smallpoly} to re-express the quantity $(2+(l-j))!(2-(l-j))!$. The Wigner 3-j symbols can be simplified by applying equation~\eqref{eq:ThreeJSymbol}.}
\begin{equation}
\gamma_l^j = (-1)^{j+l}\frac{6\sqrt{5(2j+1)(j^2+j-2)}j(j+1)(j^2+j-2-l(l+1))}{\frac{(j+l+3)!}{(j+l-2)!} \(4(l-j)^2-3|l-j|+2 \) }. \label{eq:gammalj}
\end{equation}
Since $j=0$ and $j=1$ are roots in this formula, we do not need to consider the restriction $j\geq 2$. We can now rewrite our Legendre expansion as an expansion in $\spY{2}{l}{1}$:
\begin{align*}
Q^{(1)}+\ii U^{(1)} &= -\sqrt{\frac{\pi}{5}} \sum_l (-\ii)^l (2l+1) G^{(1)}_l \sum_{j=l-2}^{l+2} \gamma_l^j \spY{2}{j}{1} \\
&= -\sqrt{\frac{\pi}{5}}  \left\{ \sum_l (-\ii)^l (2l+1) G^{(1)}_l \gamma_l^{l-2} \spY{2}{l-2}{1} + \cdots + \sum_l (-\ii)^l (2l+1) G^{(1)}_l \gamma_l^{l+2} \spY{2}{l+2}{1}
\right\} \\
&= -\sqrt{\frac{\pi}{5}}  \left\{ \sum_l (-\ii)^{l+2} (2(l+2)+1) G^{(1)}_{l+2} \gamma_{l+2}^{l} \spY{2}{l}{1} + \cdots + \right. \\
&\phantom{=-4\sqrt{\frac{\pi}{5}}} \left. +\sum_l (-\ii)^{l-2} (2(l-2)+1) G^{(1)}_{l-2} \gamma_{l-2}^{l}  \spY{2}{l}{1}  \right\}  \\
&= -\sqrt{\frac{\pi}{5}} \sum_l (-\ii)^l \left\{ \[-(2l-3)  \gamma_{l-2}^{l}G^{(1)}_{l-2} + (2l+1) \gamma_{l}^{l} G^{(1)}_l - (2l+5) \gamma_{l+2}^{l} G^{(1)}_{l+2} \] \right. \\
&\phantom{=-4\sqrt{\frac{\pi}{5}} \sum_l (-\ii)^l}
\left. +\ii \[ (2l-1)\gamma_{l-1}^l G^{(1)}_{l-1} - (2l+3)\gamma_{l+1}^l G^{(1)}_{l+1}\] \right\} \spY{2}{l}{1} .
\end{align*}
We can finally compare with Hu et al., and we find
\begin{align}
E_l^{(1)} &= \frac{1}{2}\sqrt{\frac{2l+1}{5}} \[ (2l-3) \gamma_{l-2}^l G^{(1)}_{l-2} - (2l+1) \gamma_{l}^{l} G^{(1)}_l + (2l+5) \gamma_{l+2}^{l} G^{(1)}_{l+2} \], \\
B_l^{(1)} &= \frac{1}{2}\sqrt{\frac{2l+1}{5}} \[ -(2l-1)\gamma_{l-1}^l G^{(1)}_{l-1} + (2l+3)\gamma_{l+1}^l G^{(1)}_{l+1} \] .
\end{align}
In order to calculate the $P^{(1)}$ of Hu\&White, we need $\Theta_2^{(1)}$ and $E_2^{(1)}$. They are
\begin{align}
\Theta_2^{(1)} &= -\frac{1}{4}\sqrt{3!}\(F_1^{(1)} + F_3^{(1)} \), \\
E_2^{(1)} &=\frac{1}{2} \[\gamma_0^2 G_0^{(1)}-5 \gamma_2^2 G_2^{(1)} + 9 \gamma_4^2 G_4^{(1)} \], \nonumber \\
&= \frac{1}{2} G_0^{(1)} + \frac{5}{14} G_2^{(1)} -\frac{1}{7} G_4^{(1)},
\end{align}
leading to
\begin{align}
P^{(1)} &= \frac{1}{10} \[\Theta_2^{(1)}-\sqrt{6} E_2^{(1)} \] \nonumber \\
&= -\frac{\sqrt{6}}{40} \[ F_1^{(1)} + F_3^{(1)} + 2 G_0^{(1)} + \frac{10}{7} G_2^{(1)} -\frac{4}{7} G_4^{(1)} \],
\end{align}
which matches our definition of $\mathcal{P}^{(1)}$ in equation~\eqref{eq:defPm}.

\subsection{Tensor modes}
Our expansion for $\Theta$ reads
\begin{align}
\Theta^{(2)} &= \frac{1}{4} \ee^{2\ii \phi} (1-\mu^2) F^{(2)} \nonumber \\
&= \frac{1}{4} \sum_l (-\ii)^l \ee^{2\ii \phi} (1-\mu^2) F^{(2)}_l (2l+1) P_l \nonumber \\
&= \frac{1}{4} \sum_l (-\ii)^l \ee^{2\ii \phi} (1-\mu^2) F^{(2)}_l \[ \frac{1}{2l-1} \diff{^2P_{l-2}}{\mu^2} -\frac{2(2l+1)}{(2l-1)(2l+3)} \diff{^2P_l}{\mu^2} +\frac{1}{2l+3} \diff{^2P_{l+2}}{\mu^2}\] \nonumber \\
&= \frac{1}{4} \sum_l (-\ii)^l \ee^{2\ii \phi} F^{(2)}_l \[ \frac{P_{l-2}^2}{2l-1} -\frac{2(2l+1)P_l^2}{(2l-1)(2l+3)}  +\frac{P_{l+2}^2}{2l+3}  \]\nonumber \\
&= \frac{1}{4} \sum_l (-\ii)^l \ee^{2\ii \phi} F^{(2)}_l \frac{P_{l-2}^2}{2l-1}- \sum_l (-\ii)^l \ee^{2\ii \phi} F^{(2)}_l \frac{2(2l+1)P_l^2}{(2l-1)(2l+3)} +\sum_l (-\ii)^l \ee^{2\ii \phi} F^{(2)}_l \frac{P_{l+2}^2}{2l+3}\nonumber\\
&= \frac{1}{4} \sum_l (-\ii)^l \ee^{2\ii \phi} \[-\frac{1}{2l-1} F^{(2)}_{l-2} - \frac{2(2l+1)}{(2l-1)(2l+3)} F^{(2)}_l - \frac{1}{2l+3} F^{(2)}_{l+2} \] P_l^2 \nonumber\\
&= \sum_l (-\ii)^l \frac{-1}{4} \sqrt{\frac{4\pi}{2l+1}} \sqrt{\frac{(l+2)!}{(l-2)!}} \[\frac{1}{2l-1} F^{(2)}_{l-2} + \frac{2(2l+1)}{(2l-1)(2l+3)} F^{(2)}_l + \frac{1}{2l+3} F^{(2)}_{l+2} \] Y_l^2,
\end{align}
where we used equation~\eqref{eq:Plmdef} and~\eqref{eq:Ylmdef}. This expansion can be compared directly to that of Hu\&White:
\begin{equation}
\Theta^{(2)} = \sum_l (-\ii)^l \sqrt{\frac{4\pi}{2l+1}} \Theta^{(2)}_l Y_l^2,
\end{equation}
so we find
\begin{equation}
\Theta^{(2)}_l = -\frac{1}{4} \sqrt{\frac{(l+2)!}{(l-2)!}} \[\frac{1}{2l-1} F^{(2)}_{l-2} + \frac{2(2l+1)}{(2l-1)(2l+3)} F^{(2)}_l + \frac{1}{2l+3} F^{(2)}_{l+2} \].
\end{equation}
Thus,
\begin{align}
\Theta^{(2)}_2 &= -\frac{1}{2}\sqrt{6} \[\frac{1}{3} F^{(2)}_0 + \frac{10}{21} F^{(2)}_2 + \frac{1}{7} F^{(2)}_4 \] \nonumber \\
&= -\frac{10}{\sqrt{6}} \[\frac{1}{10}F^{(2)}_0 + \frac{1}{7} F^{(2)}_2 + \frac{3}{70} F^{(2)}_4 \]. \label{eq:Theta22}
\end{align}
We can write our expansions of $Q$ and $U$ in a form which is easy to compare to the expansions of Hu and White:
\begin{align}
Q^{(2)} \pm \ii U^{(2)} &= \[1 +\frac{\spB^2}{\spE^2} \] \frac{1}{4} (1+\cos^2\theta) \ee^{2\ii\phi} G^{(2)} \nonumber \\
&=\[1+\cos^2\theta \mp \cos\theta \] \frac{1}{4} G^{(2)} \nonumber \\
&= \sum_{l=0}^{\infty} (-\ii)^l (2l+1) \sqrt{\frac{4\pi}{5}} P_l \spY{\pm 2}{2}{2} G^{(2)}_l. \label{eq:QpmiU}
\end{align}
Considering only $Q + \ii U$, we expand $P_l \spY{2}{2}{2}$ in terms of spin 2 spherical harmonics:
\begin{equation}
P_l  \spY{2}{2}{2} = \sum_j \alpha_l^j \spY{2}{j}{2},\label{eq:PlSpyExpansion}
\end{equation}
where the expansion coefficients can be found from the orthogonality of the spin-weighted spherical harmonics:
\begin{align}
\alpha_l^j &= \int \di \Omega P_l \spY{2}{2}{2} (\spY{2}{j}{2})^* \nonumber \\
&= \sqrt{\frac{4\pi}{2l+1}} \int \di \Omega \spY{0}{l}{0} \spY{2}{2}{2} \spY{-2}{2}{-2} (-1)^4 \nonumber \\
&= \sqrt{\frac{4\pi}{2l+1}} \sqrt{\frac{5 (2l+1)(2j+1)}{4\pi}} \Wthreej{l}{0}{2}{2}{j}{-2} \Wthreej{l}{0}{2}{-2}{j}{2} (-1)^{0+2-2} \nonumber \\
&= \sqrt{5(2j+1)} \Wthreej{l}{0}{2}{2}{j}{-2}^2 (-1)^{l+2+j}.
\end{align}
Using the selection rules for the Wigner 3-j symbols, we find that we only have non-zero values for $|l-2| \leq j \leq l+2$, so the sum in equation~\eqref{eq:PlSpyExpansion} contains only 5 terms. We can find a simple analytical expression for the Wigner 3-j symbol using the general expression
\begin{align}
\Wthreej{j_1}{m_1}{j_2}{m_2}{j_3}{m_3} &= (-1)^{j_1-j_2-m_3} \Delta(j_1 j_2 j_3) \prod_{k=1}^3 \sqrt{(j_k-m_k)!(j_k+m_k)!} \times \nonumber \\
                &\times \sum_s (-1)^s \frac{1}{s!(j_1+j_2-j_3-s)!(j_1-m_1-s)!(j_2+m_2-s)!} \times \nonumber \\
								&\hphantom{\times \sum_s (-1)^s} \times \frac{1}{ (j_3-j_2+m_1+s)!(j_3-j_1-m_2+s)!}, \label{eq:ThreeJSymbol}
\end{align}
where the sum is over all non-negative integers $s$ such that all the arguments of the factorials are non-negative, and $\Delta(j_1 j_2 j_3)$ is defined by
\begin{equation}
\Delta(j_1 j_2 j_3) = \sqrt{\frac{(j_1+j_2-j_3)!(j_1-j_2+j_3)!(-j_1+j_2+j_3)!}{(j_1+j_2+j_3+1)!}}.
\end{equation}
It is easy to verify that in our case, only $s=l-j+2$ is allowed. We find
\begin{align}
\Wthreej{l}{0}{2}{2}{j}{-2} &= (-1)^j 2\sqrt{6} \Delta(l2j) \sqrt{\frac{(j+2)!}{(j-2)!}} \[ (2+(l-j))! (2-(l-j))! \]^{-1} \nonumber \\
&= (-1)^j 2\sqrt{6} \sqrt{\frac{(j+2)!}{(j-2)!}} \sqrt{\frac{(l+j-2)!}{(l+j+3)!}} \frac{1}{\sqrt{(2+(l-j))!(2-(l-j))!}}.
\end{align}
Note that the fractions of factorials just select a finite number of terms which can be written explicitly. The last fraction can be written in the following way:
\begin{equation}
\chi(l-j)\equiv (2+(l-j))!(2-(l-j))! = (2+|l-j|)!(2-|l-j|)! = \left\{ \begin{array}{ll} 2!2! & |l-j|=0 \\ 3! & |l-j|=1 \\ 4! & |l-j|=2 \end{array} \right. ,
\end{equation}
which for all relevant values can be written compactly as
\begin{equation}
(2+(l-j))!(2-(l-j))! = 2\[ 4(l-j)^2 -3 |l-j| +2 \]. \label{eq:smallpoly}
\end{equation}
By combining these results we find 
\begin{equation}
\Wthreej{l}{0}{2}{2}{j}{-2} = (-1)^j 2\sqrt{6} \sqrt{\frac{(j-1)j(j+1)(j+2)}{(l+j-1)(l+j)(l+j+1)(l+j+2)(l+j+3)\chi(l-j)}},
\end{equation}
leading to
\begin{equation}
\alpha_l^j = \frac{12 \sqrt{5(2j+1)} (j-1)j(j+1)(j+2)}{(l+j-1)(l+j)(l+j+1)(l+j+2)(l+j+3)(4(l-j)^2 -3 |l-j| +2)}. \label{eq:alphalj}
\end{equation}

We can now insert the expansion~\eqref{eq:PlSpyExpansion} into equation~\eqref{eq:QpmiU}:
\begin{align*}
Q^{(2)} +\ii U^{(2)} &= \sum_{l=0}^{\infty} (-\ii)^l (2l+1) \sqrt{\frac{4\pi}{5}} G^{(2)}_l \sum_{j=l-2}^{l+2} \alpha_l^j \spY{2}{j}{2} \\
&=\sum_{l=0}^{\infty} (-\ii)^l (2l+1) \sqrt{\frac{4\pi}{5}} G^{(2)}_l \alpha_l^{l-2} \spY{2}{l-2}{2} + \\
&+\vdots\\
&+\sum_{l=0}^{\infty} (-\ii)^l (2l+1) \sqrt{\frac{4\pi}{5}} G^{(2)}_l \alpha_l^{l+2} \spY{2}{l+2}{2} \\
&=\sum_{l=0}^{\infty} (-\ii)^l \sqrt{\frac{4\pi}{5}} \spY{2}{l}{2} \sum_{j=-2}^{2} (-\ii)^j (2(l+j)+1) G^{(2)}_{l+j} \alpha_{l+j}^l \\
&=\sum_{l=0}^{\infty} (-\ii)^l \sqrt{\frac{4\pi}{5}} \spY{2}{l}{2} \bigg\{ \ii \[ (2l-1)G^{(2)}_{l-1}\alpha_{l-1}^l - (2l+3) G^{(2)}_{l+1} \alpha_{l+1}^l \] + \\
&\hphantom{=\sum_{l=0}^{\infty} (-\ii)^l} + \[-(2l-3) G^{(2)}_{l-2} \alpha_{l-2}^l + (2l+1)G^{(2)}_l \alpha_l^l -(2l+5) G^{(2)}_{l+2} \alpha_{l+2}^l \] \bigg\}
\end{align*}
This expansion can now be directly compared to Hu\&Whites expansion
\begin{equation}
Q^{(2)} + \ii U^{(2)} = \sum_l (\ii)^l \sqrt{\frac{4\pi}{2l+1}} \spY{2}{l}{2} \(E_l^{(2)} + \ii B_l^{(2)} \),
\end{equation}
leading to the identification
\begin{align}
E_l^{(2)} &= \sqrt{\frac{2l+1}{5}} \[-(2l-3) G^{(2)}_{l-2} \alpha_{l-2}^l + (2l+1)G^{(2)}_l \alpha_l^l -(2l+5) G^{(2)}_{l+2} \alpha_{l+2}^l \], \\
B_l^{(2)} &= \sqrt{\frac{2l+1}{5}} \[ (2l-1)G^{(2)}_{l-1}\alpha_{l-1}^l - (2l+3) G^{(2)}_{l+1} \alpha_{l+1}^l \].
\end{align}
Note that the square roots in $\alpha_l^j$ combine with the square root in front, so all expressions will be in terms of fractions. As an example we can calculate $E_2^{(2)}$:
\begin{align}
E_2^{(2)} &= -G^{(2)}_0 + \frac{10}{7} G^{(2)}_2 - \frac{1}{14} G^{(2)}_4 \nonumber \\
&= \frac{10}{6} \[ -\frac{3}{5} G^{(2)}_0 + \frac{6}{7} G^{(2)}_2 -\frac{3}{70} G^{(2)}_4 \]. \label{eq:E22}
\end{align}
We can now calculate $P^{(2)}$ of Hu\&White from the definition of $P^{(m)}$, equation~\eqref{eq:defPm}. $\Theta_2^{(2)}$ is given by~\eqref{eq:Theta22} and $E_2^{(2)}$ by equation~\eqref{eq:E22}. Thus
\begin{align}
P^{(2)} &= \frac{1}{10} \[ \Theta_2^{(2)} -\sqrt{6} E_2^{(2)} \] \nonumber\\
&= \frac{1}{10} \[ -\frac{10}{\sqrt{6}} \(\frac{1}{10}F^{(2)}_0 + \frac{1}{7} F^{(2)}_2 + \frac{3}{70} F^{(2)}_4 \) -\sqrt{6}  \frac{10}{6} \( -\frac{3}{5} G^{(2)}_0 + \frac{6}{7} G^{(2)}_2 -\frac{3}{70} G^{(2)}_4 \) \] \nonumber \\
&= -\frac{1}{\sqrt{6}} \[ \frac{1}{10}F^{(2)}_0 + \frac{1}{7} F^{(2)}_2 + \frac{3}{70} F^{(2)}_4 -\frac{3}{5} G^{(2)}_0 + \frac{6}{7} G^{(2)}_2 -\frac{3}{70} G^{(2)}_4 \],
\end{align}
which is the same as our definition of $\mathcal{P}^{(2)}$, equation~\eqref{eq:defPm}. This expression can be compared to the $\Psi$ used by different authors, but note that the exact expression depends on the definition of the Legendre expansion as well as the convention for the perturbation. Crittenden et. al.~\cite{Crittenden:1993ni} are using intensity fluctuation units\footnote{This can be verified by integrating their definition of the perturbed distribution function $\delta f$ over momentum.} like us, but they are omitting the $(-\ii)^l$ in their convention for Legendre expansion. Taking this into account, we find $\Psi = -\sqrt{6} \mathcal{P}^{(2)}$.

\section{Relativistic Boltzmann equation in an arbitrary gauge}\label{sec:relBoltzmann}

The relativistic Boltzmann equation is often derived in the literature in the synchronous gauge (see e.g.~\cite{weinberg2008cosmology}) or longitudinal gauge (see e.g.~\cite{Hu:1995em,Mukhanov:991646,Durrer:1127831}).
The general result is derived in~\cite{Kodama:1985bj,Peter:1208401} and quoted in~\cite{Hu:1997hp,Hu:1997mn}, but with differences in the signs of a few terms, and even with a different expression for the contribution of the metric perturbation $h_{0i}$. This justifies the presentation a full derivation in this Appendix. We stick to the metric choice
\begin{equation}
g_{\mu\nu}=a^2 (\gamma_{\mu \nu} + h_{\mu \nu})
\end{equation}
such that $x^0 \equiv \tau$ represents conformal time, and $\gamma_{00}=-1$.
In the Friedmann-Lema\^{\i}tre model, for any set of comoving coordinates, the tensor $\gamma_{\mu \nu}$ must be diagonal. Indices are raised using $\bar{g}^{\mu \nu}=\bar{g}^{\mu \rho} \bar{g}^{\nu \lambda}\bar{g}_{\rho \lambda}$ for the background, and $\delta g^{\mu \nu}=-\bar{g}^{\mu \rho} \bar{g}^{\nu \lambda} \delta g_{\rho \lambda}$ for the perturbations. For photons traveling a along a given geodesic, the four-momentum $P^{\mu}=\frac{dx^\mu}{d\lambda}$ obeys as usual to $P_\mu P^\mu=0$. Instead the four-velocity $u^{\mu}$ of a given observer is normalized to $u_\mu u^\mu=-1$. When the trajectory of an observer crosses that of a photon, the observer measures a photon energy $\omega\equiv-u_\mu P^\nu$. If the observer happens to be at rest with respect to the coordinate system, $u^i$ vanishes. In that case, the previous relations imply $u^0=1/\sqrt{-g_{00}}$ and 
\begin{equation}
\omega = -g_{0\mu} \, u^0 P^\mu =  \sqrt{-g_{00}} P^0 - a h_{0i} P^i~.
\end{equation}
It is trivial to show that at order one in perturbations, $\omega \simeq \sqrt{g_{ij}P^i P^j}$ (the relation is even exact in any gauge where $h_{0i}=0$). The direction of propagation\footnote{opposite to the direction of observation.} of a  photon is given by $n^i = P^i/\sqrt{\gamma_{ij} P^i P^j}$, such that $\gamma_{ij} n^i n^j=1$. Using $P_\mu P^\mu=0$, we see that at order zero in perturbations, $n^i$ is equal to $P^i/P^0$. The phase-space distribution of a given species can be expressed as a function of coordinates $x^\mu$, of the energy $\omega$ measured by a comoving observer, and of the direction of propagation $n^i$. In the special case of massless particles, gravitational interactions preserve the shape of the distribution, that can be expressed as:
\begin{equation}
f(x^\mu, \omega, n^i) = g\left(y(x^\mu, \omega, n^i)\right)
\end{equation}
where $g(y)$ could be a Bose-Einstein or Fermi-Dirac distribution, and the quantity $y$ involves the energy, the background temperature and the direction-dependent temperature fluctuation:
\begin{equation}
y(x^\mu, p, n^i) = \frac{\omega}{T(\tau) (1+\Theta(x^\mu, n^i))} \simeq  \frac{\omega}{T(\tau)} \left(1-\Theta(x^\mu, n^i)\right)~.
\end{equation}
The dependence of the temperature fluctuation $\Theta$ on direction appears when the tight-coupling approximation breaks down. For massless species, the collisionless Boltzmann equation (or Liouville equation) expressing the conservation of the phase-space distribution along geodesics reads
\begin{equation}
\frac{df}{d\lambda}
= g'(y) \,\, y \left[ P^0 \frac{\partial \log(y)}{\partial \tau}
+ P^i \frac{\partial \log(y)}{\partial x^i}
+ \frac{d\omega}{d\lambda} \frac{\partial \log(y)}{\partial \omega}
+ \frac{dn^i}{d\lambda} \frac{\partial \log(y)}{\partial n^i}\right] = 0~.
\end{equation}
The last  term is of order two in perturbations, since in an unperturbed universe geodesics would be straight lines, while $\Theta$ and $y$ would be isotropic. Keeping at most first-order perturbations, we get
\begin{equation}
\dot{\Theta} + \frac{\dot{T}}{T} + \frac{P^i}{P^0} \partial_i\Theta-\frac{1}{P^0}\frac{d \log \omega}{d\lambda}=0~.
\end{equation}
At order zero in perturbations we are left with
\begin{equation}
\frac{\dot{T}}{T} - \frac{1}{P^0} \frac{d \log a P^0}{d\lambda}=0~.
\end{equation}
By using the geodesic equation $\frac{d P^0}{d\lambda} = - \bar{\Gamma}^0_{\mu \nu} P^\mu P^\nu$,
one can easily show that
\begin{equation}
\frac{\dot{T}}{T} + \frac{\dot{a}}{a}=0~,
\end{equation} 
and that the temperature scales like the inverse of the scale factor. By subtracting this equation to the full one, we get the first-order equation governing temperature fluctuations:
\begin{equation}
\dot{\Theta}  + \frac{P^i}{P^0} \partial_i\Theta - \left[ \frac{1}{P^0}\frac{d \log \omega}{d\lambda} + \frac{\dot{a}}{a}\right]  =0
\end{equation}
Since $\partial_i\Theta$ is a perturbation, we can replace its coefficient $\frac{P^i}{P^0}$ by its zero-order approximation $n^i$. Also, at order one, we can use $\omega = \sqrt{g_{00}} P^0 (1-h_{0i}n^i)$, and expand the equation as:
\begin{equation}
\dot{\Theta}  + n^i \partial_i\Theta - \left[ \frac{d \log \sqrt{-g_{00}}}{d\tau} 
+ \frac{1}{P^0} \frac{d \log P^0}{d\lambda} 
- \frac{dh_{0i}}{d \tau} n^i
+ \frac{\dot{a}}{a}\right]  =0
\end{equation}
(we recall that $P^0 = d \tau / d \lambda$). 
Using the relations $g^{00}=1/g_{00}$ and $\partial_\alpha g_{\mu\nu} = \Gamma_{\alpha \mu}^\lambda g_{\lambda \nu} +  \Gamma_{\alpha \nu}^\lambda g_{\lambda \mu}$, we can express the first term between brackets as
\begin{eqnarray}
\frac{d \log \sqrt{-g_{00}}}{d\tau} 
&=& \frac{1}{2 g_{00}} \left(  \dot{g}_{00} + \partial_i g_{00} \frac{dx^i}{d\tau} \right) \nonumber\\
&=& \Gamma_{00}^0 + \left( \Gamma_{i0}^0 - \frac{\dot{a}}{a} h_{0i} \right) \frac{P^i}{P^0}~.
\end{eqnarray}
The second term between brackets can be expanded using the geodesics equation:
\begin{equation}
\frac{1}{P^0} \frac{d \log P^0}{d\lambda}  = - \Gamma^0_{00} - 2 \Gamma_{i0}^0 \frac{P^i}{P^0}
+ \Gamma_{ij}^0 \frac{P^i}{P^0} \frac{P^j}{P^0}~.
\end{equation}
After some cancellations, we are left with a reduced expression:
\begin{equation}
\dot{\Theta}  + n^i \partial_i\Theta +
\Gamma^0_{ij} \frac{P^i P^j}{(P^0)^2}
+ \Gamma^0_{0i} \frac{P^i}{P^0}
+ \frac{\dot{a}}{a} h_{0i} n^i + \frac{dh_{0i}}{d \tau} n^i
- \frac{\dot{a}}{a}  =0~.
\label{Boltzm4}
\end{equation}
The third term can be computed by substituting $\Gamma^0_{ij}$ with its explicit expression, $P^i P^j$ by $(\gamma_{\alpha \beta}P^\alpha P^\beta) n^i n^j$, and $(\gamma_{\alpha \beta}P^\alpha P^\beta)$ by $(- a^{-2} g_{00} P^0 P^0 - 2 h_{0\alpha} P^\alpha P^0 - h_{\alpha \beta} P^\alpha P^\beta)$.
After a few lines of somewhat tedious calculations, we get:
\begin{eqnarray}
\Gamma^0_{ij}  \frac{P^i P^j}{(P^0)^2} &=& \frac{\dot{a}}{a}  \left(  1 - 2 h_{0i}n^i \right) 
+ \frac{1}{2} \dot{h}_{ij} n^i n^j\nonumber \\
&+& \left(
- \frac{1}{2} \left[ \partial_j h_{0i} + \partial_i h_{0j}  \right] 
+ \frac{1}{2} \left[ h_{0i} \gamma^{ii} \partial_j \gamma_{ii} + h_{0j} \gamma^{jj} \partial_i \gamma_{jj} \right]
\right)
n^i n^j~.
\label{gamma0ij}
\end{eqnarray}
The second line can be written in a more compact and intuitive way using the covariant derivative $_{|i}$ defined for the spatial metric $\gamma_{ij}$ (in other words, based on the Christofell symbols $\gamma_{ij}^k$ computed from $\gamma_{ij}$, which are actually equal to those of the full spatial metric $g_{ij}$ computed at order zero in perturbations):
\begin{equation}
h_{0j|i} \equiv \partial_i h_{0j} - \gamma_{ij}^k h_{0k}~.
\end{equation}
We notice that 
\begin{equation}
h_{0i|j} n^i n^j = h_{0j|i} n^i n^j = \left(\partial_i h_{0j} - h_{0j} \gamma^{jj} \partial_i \gamma_{jj}\right) n^i n^j~.
\end{equation}
Hence the second line of equation~(\ref{gamma0ij}) is equal to $-h_{0i|j} n^i n^j $. The fourth term in equation~(\ref{Boltzm4}) is much easier to obtain:
\begin{equation}
\Gamma^0_{0i} \frac{P^i}{P^0} = - \frac{1}{2} \partial_i h_{00} n^i + \frac{\dot{a}}{a} h_{0i} n^i~.
\label{gamma00i}
\end{equation}
We can simplify equation~(\ref{Boltzm4}) using (\ref{gamma0ij}, \ref{gamma00i}). After several cancellations, we get:
\begin{equation}
\dot{\Theta}  + n^i \partial_i\Theta +
+ \frac{1}{2} \dot{h}_{ij} n^i n^j
- h_{0i|j} n^i n^j 
- \frac{1}{2} \partial_i h_{00} n^i  + \frac{dh_{0i}}{d \tau} n^i
=0~.
\label{Boltzm5}
\end{equation}
Several authors define the energy $\omega$ as $\sqrt{g_{00}}P^0$, neglecting the correction terms proportional to $h_{0i}$, and obtain the above equation without the last term (see e.g.~\cite{Kodama:1985bj,Peter:1208401}). If instead we take this term into account and express it as
\begin{equation}
\frac{dh_{0i}}{d \tau} n^i = \left(\dot{h}_{0i} + h_{0i|j} \frac{dx^j}{d\tau}\right)n^i = \dot{h}_{0i} n^i + h_{0i|j} n^i n^j~,
\end{equation}
we obtain as a final result
\begin{equation}
\dot{\Theta}  + n^i \partial_i\Theta 
+ \frac{1}{2} \dot{h}_{ij} n^i n^j
- \frac{1}{2} \partial_i h_{00} n^i  + \dot{h}_{0i} n^i
=0~.
\label{Boltzm6}
\end{equation}
This expression differs from its counterpart in~\cite{Hu:1997hp,Hu:1997mn} through the sign of the third and of the last terms.

\bibliographystyle{utcaps}

\bibliography{mb20paper}

\providecommand{\href}[2]{#2}\begingroup\raggedright\begin{thebibliography}{10}

\bibitem{Kaiser:1983}
N.~Kaiser, ``{Small-angle anisotropy of the microwave background radiation in
  the adiabatic theory},'' {\em MNRAS} {\bfseries 202} (1983)  1169--1180,
  \href{http://adsabs.harvard.edu/abs//1983MNRAS.202.1169K}{{\ttfamily
  adsabs:1983MNRAS.202.1169K}}.

\bibitem{Bond:1984fp}
J.~Bond and G.~Efstathiou, ``{Cosmic background radiation anisotropies in
  universes dominated by nonbaryonic dark matter},''
\href{http://dx.doi.org/10.1086/184362}{{\em Astrophys.J.} {\bfseries 285}
  (1984)  L45--L48}.

\bibitem{Crittenden:1993ni}
R.~Crittenden, J.~R. Bond, R.~L. Davis, G.~Efstathiou, and P.~J. Steinhardt,
  ``{The Imprint of gravitational waves on the cosmic microwave background},''
  \href{http://dx.doi.org/10.1103/PhysRevLett.71.324}{{\em Phys.Rev.Lett.}
  {\bfseries 71} (1993)  324--327},
\href{http://arxiv.org/abs/astro-ph/9303014}{{\ttfamily arXiv:astro-ph/9303014
  [astro-ph]}}.

\bibitem{Polnarev:1985}
{{Polnarev}, A.~G.}, ``{Polarization and Anisotropy Induced in the Microwave
  Background by Cosmological Gravitational Waves},'' {\em Soviet~Ast.}
  {\bfseries 29} (1985)  607,
  \href{http://adsabs.harvard.edu/abs//1985SvA....29..607P}{{\ttfamily
  adsabs:1985SvA....29..607P}}.

\bibitem{Kosowsky:1994cy}
A.~Kosowsky, ``{Cosmic microwave background polarization},''
  \href{http://dx.doi.org/10.1006/aphy.1996.0020}{{\em Annals Phys.} {\bfseries
  246} (1996)  49--85},
\href{http://arxiv.org/abs/astro-ph/9501045}{{\ttfamily arXiv:astro-ph/9501045
  [astro-ph]}}.

\bibitem{Seljak:1996is}
U.~Seljak and M.~Zaldarriaga, ``{A Line of sight integration approach to cosmic
  microwave background anisotropies},''
  \href{http://dx.doi.org/10.1086/177793}{{\em Astrophys.J.} {\bfseries 469}
  (1996)  437--444},
\href{http://arxiv.org/abs/astro-ph/9603033}{{\ttfamily arXiv:astro-ph/9603033
  [astro-ph]}}.

\bibitem{Zaldarriaga:1996xe}
M.~Zaldarriaga and U.~Seljak, ``{An all sky analysis of polarization in the
  microwave background},''
  \href{http://dx.doi.org/10.1103/PhysRevD.55.1830}{{\em Phys.Rev.} {\bfseries
  D55} (1997)  1830--1840},
\href{http://arxiv.org/abs/astro-ph/9609170}{{\ttfamily arXiv:astro-ph/9609170
  [astro-ph]}}.

\bibitem{Hu:1997hp}
W.~Hu and M.~J. White, ``{CMB anisotropies: Total angular momentum method},''
  \href{http://dx.doi.org/10.1103/PhysRevD.56.596}{{\em Phys.Rev.} {\bfseries
  D56} (1997)  596--615},
\href{http://arxiv.org/abs/astro-ph/9702170}{{\ttfamily arXiv:astro-ph/9702170
  [astro-ph]}}.

\bibitem{Zaldarriaga:1997va}
M.~Zaldarriaga, U.~Seljak, and E.~Bertschinger, ``{Integral solution for the
  microwave background anisotropies in nonflat universes},''
  \href{http://dx.doi.org/10.1086/305223}{{\em Astrophys.J.} {\bfseries 494}
  (1998)  491--502},
\href{http://arxiv.org/abs/astro-ph/9704265}{{\ttfamily arXiv:astro-ph/9704265
  [astro-ph]}}.

\bibitem{Hu:1997mn}
W.~Hu, U.~Seljak, M.~J. White, and M.~Zaldarriaga, ``{A complete treatment of
  CMB anisotropies in a FRW universe},''
  \href{http://dx.doi.org/10.1103/PhysRevD.57.3290}{{\em Phys.Rev.} {\bfseries
  D57} (1998)  3290--3301},
\href{http://arxiv.org/abs/astro-ph/9709066}{{\ttfamily arXiv:astro-ph/9709066
  [astro-ph]}}.

\bibitem{Ma:1995ey}
C.-P. Ma and E.~Bertschinger, ``{Cosmological perturbation theory in the
  synchronous and conformal Newtonian gauges},''
  \href{http://dx.doi.org/10.1086/176550}{{\em Astrophys.J.} {\bfseries 455}
  (1995)  7--25},
\href{http://arxiv.org/abs/astro-ph/9506072}{{\ttfamily arXiv:astro-ph/9506072
  [astro-ph]}}.

\bibitem{Challinor:1998xk}
A.~Challinor and A.~Lasenby, ``{Cosmic microwave background anisotropies in the
  CDM model: A Covariant and gauge invariant approach},''
  \href{http://dx.doi.org/10.1086/306841}{{\em Astrophys.J.} {\bfseries 513}
  (1999)  1--22},
\href{http://arxiv.org/abs/astro-ph/9804301}{{\ttfamily arXiv:astro-ph/9804301
  [astro-ph]}}.

\bibitem{Challinor:1999xz}
A.~Challinor, ``{Microwave background anisotropies from gravitational waves:
  The (1+3) covariant approach},''
  \href{http://dx.doi.org/10.1088/0264-9381/17/4/309}{{\em Class.Quant.Grav.}
  {\bfseries 17} (2000)  871--889},
\href{http://arxiv.org/abs/astro-ph/9906474}{{\ttfamily arXiv:astro-ph/9906474
  [astro-ph]}}.

\bibitem{Challinor:2000as}
A.~Challinor, ``{Microwave background polarization in cosmological models},''
  \href{http://dx.doi.org/10.1103/PhysRevD.62.043004}{{\em Phys.Rev.}
  {\bfseries D62} (2000)  043004},
\href{http://arxiv.org/abs/astro-ph/9911481}{{\ttfamily arXiv:astro-ph/9911481
  [astro-ph]}}.

\bibitem{Lewis:2002nc}
A.~Lewis and A.~Challinor, ``{Evolution of cosmological dark matter
  perturbations},'' \href{http://dx.doi.org/10.1103/PhysRevD.66.023531}{{\em
  Phys.Rev.} {\bfseries D66} (2002)  023531},
\href{http://arxiv.org/abs/astro-ph/0203507}{{\ttfamily arXiv:astro-ph/0203507
  [astro-ph]}}.

\bibitem{Lewis:2004ef}
A.~Lewis, ``{CMB anisotropies from primordial inhomogeneous magnetic fields},''
  \href{http://dx.doi.org/10.1103/PhysRevD.70.043011}{{\em Phys.Rev.}
  {\bfseries D70} (2004)  043011},
\href{http://arxiv.org/abs/astro-ph/0406096}{{\ttfamily arXiv:astro-ph/0406096
  [astro-ph]}}.

\bibitem{weinberg2008cosmology}
S.~Weinberg, {\em Cosmology}.
\newblock OUP Oxford, 2008.

\bibitem{Abbott:1986ct}
L.~Abbott and R.~K. Schaefer, ``{A general, gauge invariant analysis of the
  cosmic microwave anisotropy},''
\href{http://dx.doi.org/10.1086/164525}{{\em Astrophys.J.} {\bfseries 308}
  (1986)  546}.

\bibitem{Hu:1995em}
W.~T. Hu, ``{Wandering in the Background: A CMB Explorer},''
\href{http://arxiv.org/abs/astro-ph/9508126}{{\ttfamily arXiv:astro-ph/9508126
  [astro-ph]}}.

\bibitem{Mukhanov:991646}
V.~Mukhanov, {\em Physical Foundations of Cosmology}.
\newblock Cambridge Univ. Press, Cambridge, 2005.

\bibitem{Durrer:1127831}
R.~Durrer, {\em The Cosmic Microwave Background}.
\newblock Cambridge Univ. Press, Cambridge, 2008.

\bibitem{Kodama:1985bj}
H.~Kodama and M.~Sasaki, ``{Cosmological Perturbation Theory},''
\href{http://dx.doi.org/10.1143/PTPS.78.1}{{\em Prog.Theor.Phys.Suppl.}
  {\bfseries 78} (1984)  1--166}.

\bibitem{Peter:1208401}
P.~Peter and J.-P. Uzan, {\em Primordial cosmology}.
\newblock Oxford Graduate Texts. Oxford Univ. Press, Oxford, 2009.

\end{thebibliography}\endgroup

\end{document}